\documentclass{article}
\usepackage[accepted]{icml2022}
\usepackage{fancyhdr}
\usepackage{framed,multirow}
\usepackage{natbib}
\usepackage{graphicx}
\usepackage{amssymb}
\usepackage{latexsym}

\usepackage{url}
\usepackage{xcolor}

\usepackage{algorithmic}
\usepackage{booktabs} 
\usepackage[colorlinks=false,linkcolor=blue]{hyperref}
\usepackage{bbm}
\usepackage{amsmath,amsthm,amsfonts,amssymb,amscd}
\usepackage{csvsimple}
\usepackage{xcolor}
\usepackage[normalem]{ulem} %
\usepackage{array}

\usepackage{csquotes}

\usepackage{soul}
\definecolor{black}{RGB}{0,0,0}
\definecolor{blue}{RGB}{0,0,0}
\definecolor{orange}{RGB}{0,0,0}

\usepackage{makecell} 
\newcommand{\PreserveBackslash}[1]{\let\temp=\\#1\let\\=\temp}
\newcolumntype{C}[1]{>{\PreserveBackslash\centering}p{#1}}
\newcolumntype{R}[1]{>{\PreserveBackslash\raggedleft}p{#1}}
\newcolumntype{L}[1]{>{\PreserveBackslash\raggedright}p{#1}}
\definecolor{newcolor}{rgb}{.8,.349,.1}
\usepackage{placeins}

\title{\textbf{Invisible Users: Uncovering End-Users' Requirements for Explainable AI via Explanation Forms and Goals}}

\author{Weina Jin$^1$ \and Jianyu Fan$^2$ \and Diane Gromala$^2$ \and Philippe Pasquier$^2$ \and Ghassan Hamarneh$^1$}
\date{%
    $^1$School of Computing Science, Simon Fraser University\\%
    $^2$School of Interactive Arts and Technology, Simon Fraser University\\
    [2ex]%
}

\begin{document}
\maketitle
\footnotetext[1]{\copyright 2023. This manuscript version is made available under the CC-BY-NC-ND 4.0 license \href{https://creativecommons.org/licenses/by-nc-nd/4.0/}{https://creativecommons.org/licenses/by-nc-nd/4.0/}}

\begin{abstract}
    Non-technical end-users are silent and invisible users of the state-of-the-art explainable artificial intelligence (XAI) technologies. Their demands and requirements for AI explainability are not incorporated into the design and evaluation of XAI techniques, which are developed to explain the rationales of AI decisions to end-users and assist their critical decisions. This makes XAI techniques ineffective or even harmful in high-stakes applications, such as healthcare, criminal justice, finance, and autonomous driving systems. To systematically understand end-users' requirements to support the technical development of XAI, we conducted the EUCA user study with 32 layperson participants in four AI-assisted critical tasks. The study identified comprehensive user requirements for feature-, example-, and rule-based XAI techniques (manifested by the end-user-friendly explanation forms) and XAI evaluation objectives (manifested by the explanation goals), which were shown to be helpful to directly inspire the proposal of new XAI algorithms and evaluation metrics. The EUCA study findings, the identified explanation forms and goals for technical specification, and the EUCA study dataset support the design and evaluation of end-user-centered XAI techniques for accessible, safe, and accountable AI.

\end{abstract}

\section{Introduction}
Doctors, judges, drivers, bankers, and other decision-makers may request explanations from artificial intelligence (AI) when using it to support their critical decisions, such as identifying disease~\cite{Caruana2015,Jin2020,Holzinger2017a}, reaching a guilt verdict~\cite{Kleinberg2017}, ensuring safety of an autonomous driving vehicle, and approving loan applications.
As AI becoming pervasive as decision assistant in high-stakes domains -- such as medicine, criminal justice, finance, and autonomous driving systems --making AI explainable to its users is crucial to ensure the safe, responsible, and legal use of AI~\cite{gdpr}. This is the interpretable or explainable AI (XAI) problem~\cite{Doshi-Velez2017}. The XAI technical development, however, mainly focuses on addressing technical users' demand for AI explanation for model debugging, understanding, and improvement. This research paradigm disproportionately ignores the demand for XAI from the largest and most diverse group of users: the non-technical end-users~\cite{Miller, Bhatt2019,10.1145/3411764.3445088,Laato2022}, or end-users for short. Previous user studies have shown that, directly reorienting existing XAI techniques that are originally designed for technical users may not comply with end-users' reasoning process~\cite{JIN2023102684, Jin_Li_Hamarneh_2022,rethinking}, may not fulfill end-users' demands such as verifying model decisions~\cite{NEURIPS2021_de043a5e, shen2020useful,Kim2022HIVE} and detecting model biases~\cite{adebayo2022post}, and may even lead to harmful consequences such as worsening of physicians' task performance~\cite{10.1145/3375627.3375833,Jacobs2021}.
In the design and evaluation of XAI techniques, developers and researchers tend to simplify or even ignore end-users' needs, assuming that end-users' requirements for XAI are the same as technical users'. This makes end-users the invisible and silent users of XAI, whose requirements are not taken into account to shape the technologies that assist them. Without thorough understandings of end-users' requirements for XAI, it is difficult to develop XAI techniques -- including XAI algorithms and user interfaces -- that can meet their needs for AI explainability and are useful for end-users' critical tasks~\cite{9920137,Chazette2022}.

Understanding end-users' requirements for XAI, however, faces its unique challenges. First, different from technical users, 
end-users do not have technical knowledge in AI, data science, or programming. This makes it difficult to directly communicate with end-users about their requirements for XAI techniques. 
Second, unlike technical users' demand for XAI that focuses on a unified requirement to debug and improve AI models~\cite{Hong2020,Ferreira2020}, 
end-users have diverse roles, tasks, and demands for XAI.
This makes it challenging to know whether the improvement based on a particular user requirement can meet their diverse needs, and if so, how to evaluate for it.

To design and evaluate XAI techniques for end-users' purposes, in this work, we aim to gain a deep understanding of the complex nature of user requirements by conducting the EUCA (\underline{E}nd-\underline{U}ser-\underline{C}entered explainable \underline{A}I) user study. In the study design, to tackle the first challenge and overcome the technical communication barrier between XAI techniques and end-users, we propose the \textbf{end-user-friendly explanation forms}. Explanation form is the abstracted structure of explanatory reasoning information generated by an XAI algorithm. An end-user-friendly explanation form has an end-user-facing property of its various visual representations, and an AI designer-facing property of its different underlying XAI algorithms. Thus, the explanation form is a bridging interface designed to categorize the XAI algorithm space according to end-user-centered properties, as represented by the forms of feature- (feature attribution, feature shape, feature interaction), example- (similar, typical, and counterfactual example), and rule-based explanations (decision rule and decision tree). Explanation forms ``translate'' XAI techniques to users to acquire their comments and requirements, and users' requirements in turn can be ``translated back'' as technical specifications to directly motivate the proposal of new XAI techniques, as shown in the discussion Section~\ref{form_design}.
To tackle the second challenge of users' various demands for XAI, we summarize from literature the main motivations for end-users to check explanations as 10 \textbf{explanation goals}, including to calibrate trust, ensure safety, detect bias, resolve disagreement between user and AI,
differentiate similar instances, learn from AI, improve user's predicted outcome, communicate with stakeholders, generate reports, and trade off multiple objectives. Understanding users' requirements for explanation goals can help to formulate new human-subject or computational evaluation objectives on XAI utility, as discussed in Section~\ref{goal_eval}.  

To understand end-users' requirements for XAI techniques (with respect to their explanation forms) and explanation goals, we conducted a mixed-method user study with 32 layperson participants. The user study asked participants to select, rank, and comment on explanation forms to fulfill an explanation goal in AI-supported critical tasks, including buying a house, using autonomous driving vehicle, checking the health risk of diabetes, and preparing for an exam. The study findings on user requirements for each explanation form and goal are summarized in Table~\ref{tab:result} and~\ref{tab:rq2}. We further discuss with examples on how the understandings of users' requirements for explanation forms and goals can support the design and evaluation of end-user-centered XAI techniques.

Our contributions are:
\textbf{1}. We conduct the first comprehensive user study to systematically identify end-user requirements for XAI techniques. The EUCA study covers 12 explanation forms, 10 explanation goals, and interactions between the two. 
It has the advantage that users' requirements are acquired through comparisons among different explanation forms and in the context of end-user-oriented explanation goals.

\textbf{2}. We propose the end-user-friendly explanation forms. We also identify the explanation forms and explanation goals as effective tools to understand end-users' requirements for XAI techniques. Understandings of users requirements for \textbf{explaination forms} can directly inspire the \textbf{design} of new end-user-centered XAI techniques; and understandings of users requirements for \textbf{explaination goals} can support the proposal of new XAI \textbf{evaluation} objectives, methods, and computational metrics to assess the utility of XAI techniques to end-users.

\textbf{3}. We have made the EUCA dataset (quantitative and qualitative data of the user study), the detailed qualitative results, and the study design materials publicly available at the EUCA website (\href{https://weinajin.github.io/euca}{https://weinajin.github.io/euca}) to support the design and evaluation of end-user-centered XAI techniques in the community.

\begin{table*}[!h]
    \caption{\textbf{Key Findings on RQ1: 
  End-Users' Requirements for Explanation Forms}. For an explanation form, we also list its typical visual representations and XAI algorithms. $*$ indicates the requirement has inspired the proposal of new XAI algorithm in~\cite{transcending}, discussed in detail in Section~\ref{form_design}. The same for Table~\ref{tab:rq2}.
    }
    \centering
    \small
    \begin{tabular}{L{2cm}L{7cm}L{1.5cm}L{5cm}}
    \toprule
     \textbf{Explanation Form}    &  \textbf{End-Users' Interpretation \& Requirement} & \textbf{Visual Representation} & \textbf{XAI Algorithm}  \\
     \toprule
      \textbf{Feature attribution}   & 
      Can answer how and why AI reaches its decisions;
      Can assess AI by comparing features with user's prior knowledge$^*$; 
     Users may require different levels of detail on the attribution & 
     Saliency map; Bar chart 
     & 
     Linear or scoring models~\citep{DBLP:journals/ml/UstunR16}, Shapley value~\citep{RM-670-PR}, SHAP~\citep{NIPS2017_8a20a862}, TCAV~\citep{Kim}, LIME \cite{Ribeiro2016a}, CAM \cite{Zhoub}
     \\
      \hline
    \textbf{Feature shape} &  Graphic representation is easy to understand; Support counterfactual reasoning & Line plot & 
    GAM~\cite{Tan}, PDP~\cite{Friedman2001}, ALE~\cite{Apley2016}, SHAP dependence~\cite{Lundberg}\\
      \hline
     \textbf{Feature interaction} & Difficult to interpret; Require to prioritize significant feature interactions & 2D or 3D heatmap & 
     GA2M~\cite{Caruana2015}, PDP~\cite{Friedman2001}, ALE~\cite{Apley2016}, SHAP interaction~\cite{Lundberg} \\ 
     \toprule
     \textbf{Similar example} & Similar to human decision-making; Require to support side-by-side feature-based comparison among examples & Data instance & 
     $k$-nearest neighbors, graph-based approach~\citep{Kawaharaa}, CBR~\citep{Kolodner1992}
     \\  \hline
     \textbf{Typical example} & May be misinterpreted as similar example; Require atypical examples to show edge cases for safety and biases check$^*$ & Data instance & 
     $k$-mediods, MMD-critic \cite{Kim2016}, CNN prototype~\cite{Lia, Chen,Simonyan, Mahendran2014a}
     \\ 
     \hline
     \textbf{Counterfactual example} & Can suggest improvements, and help users differentiate similar instances; User can define the counterfactual outcome, the counterfactual feature type and range$^*$ & Data instance
     & 
     Visual counterfactual~\citep{pmlr-v97-goyal19a}, pertinent negatives~\citep{Dhurandhar}
\\ 
     \toprule
     \textbf{Decision rule} & Decision logic is simple and understandable; Require to balance explanation completeness and simplicity$^*$ &  Text, table, or matrix & Bayesian rule lists~\citep{Yang2017}, LORE~\citep{Guidotti2018c}, Anchors~\citep{Ribeiro}\\ \hline
     \textbf{Decision tree} &   Difficult to interpret; Can support counterfactual reasoning; Require to highlight branches for user's interested instances &Tree diagram & Disentangled CNNs \citep{Zhang2018d}; model distillation \citep{Frosst} \\
    \bottomrule
    \end{tabular}
    \label{tab:result}
\end{table*}

\begin{table*}[!h]
  \caption{
  \textbf{Key Findings on RQ2: 
  End-Users' Requirements for Explanation Goals}.
  }
  \label{tab:rq2}
  \renewcommand{\arraystretch}{1.2} %
\small
  \begin{tabular}{L{4cm} L{12cm}}
    \toprule
    \textbf{Explanation Goal} & \textbf{End-Users' Requirement} \\
    \toprule
    \makecell[tl]{\textbf{\textcolor{black}{Calibrate trust}} } & Require information on AI model performance, important features$^*$, the capability and credit of AI.
    \\
    \hline
    \makecell[tl]{\textbf{\textcolor{black}{Ensure safety}} } &\makecell[tl]{
    Performance in various testing cases to show AI's robustness on safety.}
    \\
    \hline
    \makecell[tl]{\textbf{\textcolor{black}{Detect bias}} }&
    \makecell[tl]{Require AI to maintain the same capability and
    perform well for minority subgroups.}
    \\
    \hline
    \textbf{Users disagree with AI}& 
    Users may lose trust; Need explanations for decision verification; User may check input for debugging.
\\
    \hline
    \textbf{Users agree with AI} & 
    Users may not need explanations; Explanations are used to boost decision confidence and improve user's outcome.
    \\
    \hline    
\textbf{Differentiate similar instances} &
    Be able to discern similar instances and pinpoint the feature differences.
    \\
    \hline
    \textbf{Learn from AI} &
    Depend on how reliable and trustworthy the AI is;
    Require a wide range of explanation to support user's learning.
    \\
    \hline

    \textbf{Improve user's outcome} &
    Seek explanation with controllable features.
    \\
    \hline

    \makecell[tl]{\textbf{\textcolor{black}{Communicate}}}
    &
    Other stakeholders need to have a basic understanding of AI; Require a formal summary or report from AI.
    \\
    \hline
    \makecell[tl]{\textbf{\textcolor{black}{Generate reports}} } &
    \makecell[tl]{
    Require a summary of key factors that lead to the decision.
    }
    \\
    \hline
    \textbf{Multi-objectives trade-off} &
    Allow users to take over; Users use explanations to defend for or against certain objectives.
    \\
  \bottomrule
\end{tabular}
\end{table*}

\section{Related work}

\textbf{Human-centered XAI and user study}
The human-computer interaction (HCI) community calls for human-centered XAI to propose and assess XAI techniques based on human-centered properties ~\citep{DBLP:journals/corr/abs-2110-10790, 10.7551/mitpress/12186.003.0014}.
User studies were conducted to understand laypersons' or domain experts' perceptions of different explanation forms in various domains and tasks, including feature attribution~\cite{EVANS2022281}, similar example~\citep{Cai2019, 10.1145/3290605.3300234}, typical example~\cite{EVANS2022281}, counterfactual example~\cite{Cai2019, 10.1145/3531146.3533189,EVANS2022281}, decision rule~\cite{10.1016/j.dss.2010.12.003}, decision sets~\citep{Narayanan2018, 10.1016/j.dss.2010.12.003}.
User study is also an important method to evaluate the user-centered utility of XAI techniques, including the direct and indirect utility for users to use explanation. User studies on the direct utility of XAI assess how effectively end-users can use the AI explanation to answer their inquiries that motivate them to seek explanations (we call them explanation goals), such as trust calibration~\cite{Zhang2020a,7349687,10.1145/3375627.3375833, 10.1145/3377325.3377480}, decision verification~\cite{NEURIPS2021_de043a5e, shen2020useful,Kim2022HIVE}, and bias detection~\cite{adebayo2022post}. Users' direct inquiry of explanations is usually an intermediate step in users' entire decision-making process, and previous user studies also evaluated the indirect utility of XAI to users' decision-making, such as human-AI team performance~\cite{Jin2022.12.07.22282726, taesiri2022visual, siu2021evaluation}.
Despite the research advances on user study, 
the technical adoption to user-centered XAI is relatively slow. Our work joins prior user studies on human-centered XAI, and differs by our unique study design to include a comprehensive set of explanation forms and goals. This enables users to directly compare all available explanation forms in the same setting and in the context of different explanation goals. We also differ from prior works by our identification of end-user-friendly explanation forms that enable convenient translation from user study findings to XAI technical development.

\textbf{XAI and information visualization}
XAI is an active research topic in information visualization (VIS) community. However, current VIS research is mainly motivated by technical users' demands of XAI for model monitoring and debugging~\citep{ALICIOGLU2022502}. To design for the large neglected group of end-users, \citet{Ming2019} proposed an interactive XAI visualization RuleMatrix for end-users to understand, explore, and validate black-box AI models. \citet{Jin2019vis} proposed the end-user-centered XAI visual vocabularies, which summarize the visual representations of XAI algorithms to bridge AI developers with end-users. Their works inspired ours on the visual representations of explanation forms.

\section{End-user-friendly explanation forms}\label{ef}
\begin{figure*}[h]
    \centering
    \includegraphics[width =\linewidth]{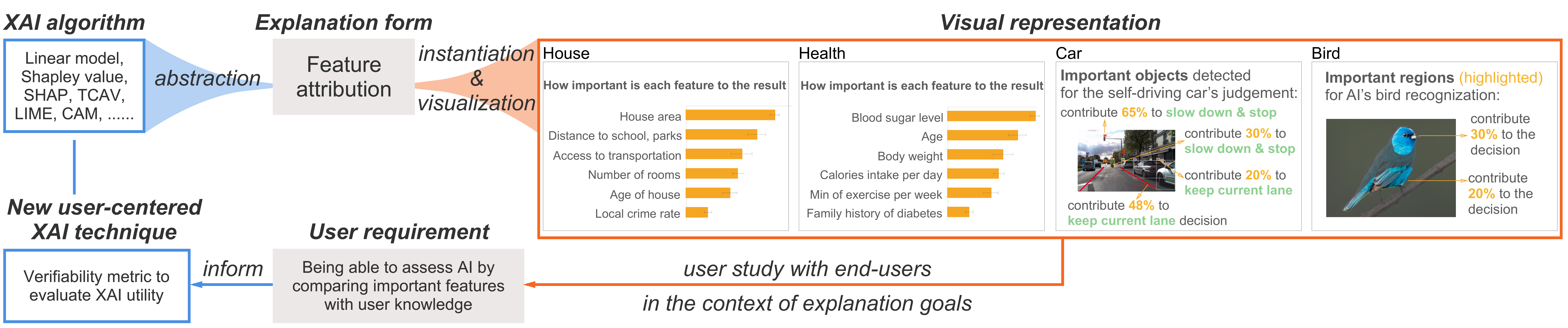}
    \caption{The process of using explanation forms to inform user-centered XAI techniques, using the explanation form of feature attribution as an example: feature attribution is abstracted from XAI algorithms and instantiated for four tasks in the user study, then a particular user requirement for feature attribution informs the proposal of new XAI techniques -- the verifiability metric (Section~\ref{goal_eval}). 
    }
    \label{fig:fa}
\end{figure*}

We propose the end-user-friendly explanation forms as a communication tool to acquire end-users' requirements for XAI techniques. As shown in Fig.~\ref{fig:fa}, we abstract an XAI algorithm by extracting and categorizing its distinctive explanatory reasoning process as an explanation form, regardless of its computational process, the particular explanatory content it generates, the data and model type, and the specific domain and task. In the user study, we then instantiate an explanation form as its visual representations by filling the form with explanatory contents to tailor to a specific data and task. 
An explanation form thus has its end-user-oriented visual interface and engineer-oriented backend XAI algorithms, and the abstraction by its form and the instantiation of the form as visual representations are two key elements for its translational role between end-users and the XAI techniques.
To identify the end-user-friendly explanation forms, we conduct a literature review and examine XAI techniques and surveys in the AI, VIS, and HCI domains, extract the output explanation forms from XAI algorithms, and filter out forms that require technical background to interpret (such as visualization of artificial neurons~\cite{Olah2017}). Next, we introduce the definition of each explanation form, with examples of its XAI algorithm and visual representation in Table~\ref{tab:result}. The literature review process and the figures of visual representations are in the Appendix. 

\subsection{Feature-based explanations}
(1) \textbf{Feature attribution} decomposes the to-be-explained outcome $y$
into a set of important features $t_i$,
each having a marginal attribution $a_i$ to $y$, $y=\sum_{i} a_i t_i$.

(2) \textbf{Feature shape} describes the marginal effect of a feature $t_i$ to the outcome $y$: $f_y(t_i)$, where $f$ is the function that describes $y$.

(3) \textbf{Feature interaction} is an extension of feature shape by describing the joint marginal effect of more than one feature to the outcome: $f_y(t_i, t_j)$.

\subsection{Example-based explanations}
(4) \textbf{Similar example} is an instance $x_s$ in the input space $X$ that shares similar features with the query input $x$ regardless of their predictions. Mathematically, similar examples are instances that have minimal distance to the query: $\arg \min_{x_s \in X} d(x_s, x)$, where $d$ is the distance or dissimilarity measure.

(5) \textbf{Typical example} or prototypical example is a representative instance of a prediction class, i.e.: an instance $x_t$ that maximizes the probability of the target prediction $y_t$: $\arg\max_{x_t \in X} p(f(x_t)=y_t)$, where $p$ denotes the probability, and $f$ denotes the AI model. 

(6) \textbf{Counterfactual example} is an instance $x_c$ whose features are minimally dissimilar to those of the query input $x$, yet its outcome $y_c$ is different from the target outcome $y_t$ of $x$, i.e.: $\arg\min_{x_c \in X} d(x_c, x), s.t. \max p(f(x_c)=y_c)), y_c \neq y_t$. 

\subsection{Rule-based explanations}
(7) \textbf{Decision rules and decision sets} are IF-THEN statements with conditions and predictions. 

(8) \textbf{Decision tree} represents rules graphically using a tree structure, with branches representing the decision pathways, and leaves representing the predicted outcomes. 

\subsection{Contextual information} 
To provide necessary context and background for a more complete explanation, we additionally include contextual information in the end-user-friendly explanation forms, include:

(9) \textbf{Input} $x$. 

(10) \textbf{Output} $y$.

(11) \textbf{Performance}: Model's performance metrics (such as prediction accuracy, confusion matrix, ROC, mean squared error) help end-users to judge a model's overall decision quality and set a proper expectation on model's capability.

(12) \textbf{Dataset}: It is the proper description of the training and validation dataset, such as data distribution, and how the data were collected.

\section{End-user-oriented explanation goals}\label{purpose}
We summarize from literature the following end-user-oriented explanation goals, i.e.: user's trigger point or motivation to check the explanation of an AI system.

    \noindent $\bullet$ \textcolor{blue}{\textbf{Calibrate trust}}: trust is key to establish human-AI decision-making partnership. Since users can easily distrust or overtrust AI,  it is important to calibrate trust to reflect the capabilities of AI systems ~\cite{Turner,Zhang2020}.
    
    \noindent $\bullet$ \textcolor{blue}{\textbf{Ensure safety}}: users need to ensure safety of the decision consequences~\cite{Doshi-Velez2017}.
    
    \noindent $\bullet$ \textcolor{blue}{\textbf{Detect bias}}: users need to ensure the decision is impartial and unbiased~\cite{Lim2019, Nunes2017}.
    
    \noindent $\bullet$ \textcolor{blue}{\textbf{Unexpected prediction}}: the AI prediction is unexpected, and/or users disagree with AI's prediction~\cite{Gregor1999a}.
    
    \noindent $\bullet$ \textcolor{blue}{\textbf{Expected prediction}}: AI's prediction aligns with users' expectations~\cite{Gregor1999a}.
    
    \noindent $\bullet$ \textcolor{blue}{\textbf{Differentiate similar instances}}: due to the consequences of wrong decisions, users sometimes need to discern similar instances or outcomes. For example, a doctor differentiates whether the diagnosis is a benign or malignant tumor~\cite{Lim2019}.
    
    \noindent $\bullet$ \textcolor{blue}{\textbf{Learn from AI}}: users need to gain knowledge, improve their problem-solving skills, and discover new knowledge~\cite{Gregor1999a, Doshi-Velez2017, Lim2019, Nunes2017}.
    
    \noindent $\bullet$ \textcolor{blue}{\textbf{Improve the predicted outcome}}: users seek causal factors to control and improve the predicted outcome~\cite{Miller2017, Lim2019, Nunes2017}.
    \noindent $\bullet$ \textcolor{blue}{\textbf{Communicate with stakeholders}}: many critical decision-making processes involve multiple stakeholders, and users need to discuss the decision with them~\cite{Miller2017}.
    
    \noindent $\bullet$  \textcolor{blue}{\textbf{Generate reports}}: users need to utilize the explanations to perform particular tasks such as report production. For example, a radiologist generates a medical report on a patient's X-ray image~\cite{Gregor1999a}.
    
    \noindent $\bullet$ \textcolor{blue}{\textbf{Multiple objectives trade-off}}: AI may be optimized on an incomplete objective while users seek to fulfill multiple objectives in real-world applications. For example, a doctor needs to ensure a treatment plan is effective as well as having acceptable patient adherence. Ethical and legal requirements may also be regarded as objectives~\cite{Doshi-Velez2017}.

\begin{figure*}[!h]
 \centering 
\includegraphics[width=1.0\textwidth]{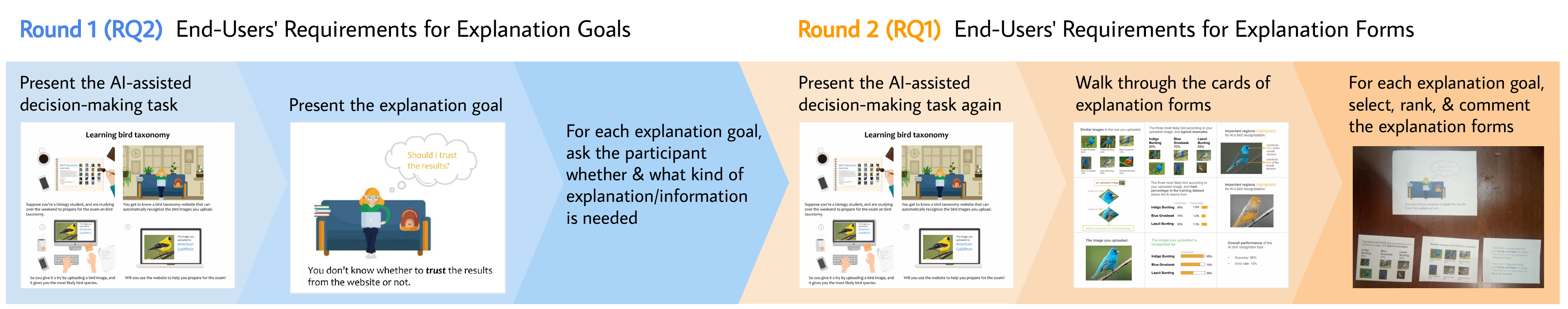}
 \caption{\textbf{The user study procedure}. The study consisted of two rounds for the two research questions. We use the Bird task and the explanation goal of calibrating trust as an example to illustrate the study procedure.}
 \label{fig:procedure}
\end{figure*}
\section{User study method}
\textbf{Study design} The EUCA user study aims to answer two research questions (RQ): What are end-users' requirements for each explanation form (RQ1), and for each explanation goal (RQ2)?
We use scenario-based requirement elicitation method~\cite{10.5555/772072.772137,Wolf2019,scenario,fraud} and include four AI-assisted critical tasks:
\textit{House} -- users use AI to estimate their house price; \textit{Health} --  users use AI to predict their diabetes risk; \textit{Car} -- users decide whether to buy an autonomous driving vehicle; and \textit{Bird} -- users use an AI bird recognition tool to prepare for an important biology exam. The four tasks  cover various input data types of AI models including tabular, sequential, image, and video data, respectively. We adapt applicable explanation goals to each task, and create a set of visual representation cards that cover all applicable explanation forms for each task. The tasks and explanation goals are presented as paper storyboards with graphics and text (Fig.~\ref{fig:procedure}).

\textbf{Study procedure} The study is an in-person, one-to-one, one-session, open-ended, semi-structured interview. It consists of two rounds (Fig.~\ref{fig:procedure}). After obtaining the participant's written informed consent, s/he enters Round 1. The participant is introduced to 
a task and a set of its explanation goals randomly. 
We ask the participant what information s/he requires for an explanation goal. 
After that, the participant enters Round 2. We first give a 5-min tutorial on the set of visual representation cards of explanation forms. The participant could ask questions if s/he did not understand or needed clarification. S/he could also comment on each card. We confirm participants' understanding of the explanation forms before moving on to the next step. 
We then ask the participant to select and rank one or multiple explanation form cards for a given explanation goal in a task, and their reasons for choosing or discarding an explanation form. The participant can also modify the given cards, sketch on blank cards to create new cards, and add the new cards to the
card ranking. We finally ask whether the chosen cards would fulfill the explanation goal.
At the end of the interview, the participants fill out a demographic questionnaire. The mean and median study duration were both 67 minutes. We audio-record the interviews, make observational notes on the card selection \& ranking process, and take pictures of the card selection \& ranking results.

\textbf{Participants and Recruitment} After ethics approval from the university's Research Ethics Board (Ethics number: 2019s0244), we recruited 32 adult participants (P1-P32) who do not have prior technical knowledge in machine learning, artificial intelligence, or data science. 
Participants were recruited via a convenience sampling method by advertising posters at public libraries, community centers, and online community boards in the metropolitan area over a 3-month period in 2019. 
Participants' demographics were: female:male = 1:1; age range 19-73, age mean$\pm$std: 38.2$\pm$16.0. Participants' occupations covered a variety of industries including technology, design, car insurance, finance, psychology, construction, sales, food/cooking, law, healthcare, government/social services, and retired. For participants who use AI in work or life (6 participants, 19\%), they used AI software such as Google Assistant to play music, navigate traffic, chat with clients, and help drive investment decisions. Participants were thanked with \$25 CAD for their time and effort in the study. Details on participants' information, study method, and study material are in the Appendix. 

\textbf{Data analysis}
We performed qualitative data analysis informed by thematic analysis~\citep{Braun2012} on the transcripts of 2,175 minutes ($>$36 hrs) of audio recordings.
We first set a priori themes and codes of each explanation form and goal based on our two research questions. With two rounds of coding and discussion on the first four transcripts, the first and second author reached a unified coding scheme and an inter-rater reliability Kappa score of 0.88. Then each author coded half of the transcripts.
We also performed quantitative data analysis on the 248 sets of data on participants' selection of the explanation forms for each explanation goal.

\begin{figure*}[!ht]
    \centering
    \includegraphics[width=\linewidth]{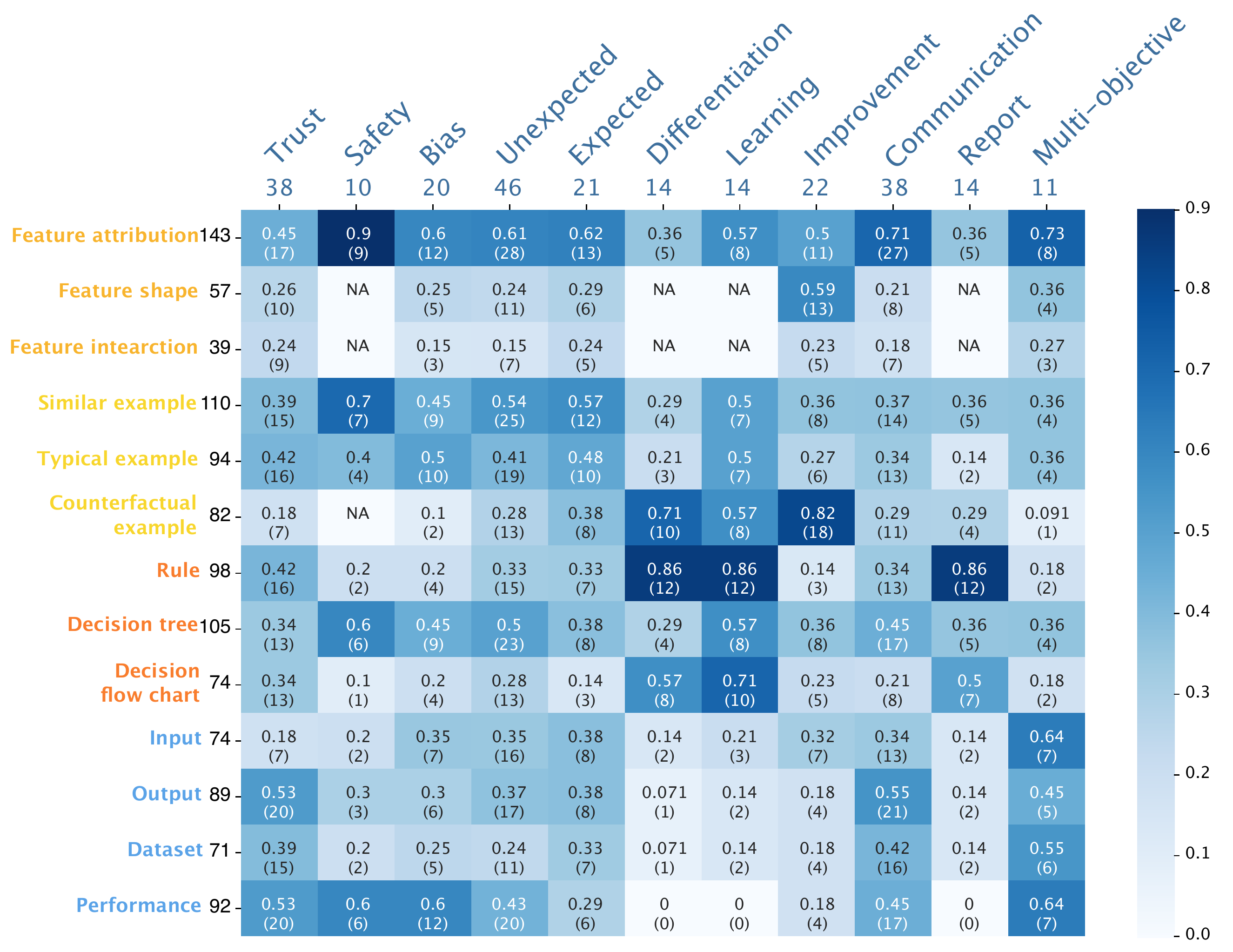}
    \caption{
    Quantitative results on the occurrence rate (first line in a cell, also color-coded by the blueness of a cell) and the number of occurrence (second line in a cell) an explanation form (row) was selected 
    for an explanation goal (column). 
    The number beside an explanation form or goal indicates the total number of an explanation form or goal among the 248 sets of data. ``NA'' indicates that an explanation form is not  available for an explanation goal.
    We include decision flow chart, which is a visual representation variation of decision tree, as a standalone explanation form. 
    }
    \label{fig:mtx}
\end{figure*}

\section{RQ1: End-users' requirements for explanation forms}
The quantitative results 
on explanation forms selected for different explanation goals are shown in Fig.~\ref{fig:mtx}. It shows that different explanation forms have different patterns for their suitable explanation goals. For example, feature attribution has a wide suitability for many explanation goals, with its occurrence rate above $0.5$ for 72\% (8/11) of the explanation goals;
while others, like counterfactual example and rule, are only appropriate for a few explanation goals.
Next, we present the qualitative results on each explanation form, which show detailed contexts and reasons on their suitability for different \ul{explanation goals} (indicated by \ul{underline}), users' interpretation, and users' requirements for improvment.
\textbf{Key findings} are summarized in Table~\ref{tab:result} and are highlighted \textbf{in bold}.

\subsection{Feature-based explanation}
\subsubsection*{(1) Feature attribution}
Participants \textbf{intuitively understand} the displayed \textit{feature} information and prefer the \textit{``simple way to highlight the most important parts''} (P04, for saliency map), and \textit{``outline of everything (features)''} (P10, for feature bar plot). 
By showing \textit{``finer details''} (P10) and \textit{``breakdown and weights of features''}(P23) \textit{``that AI took into account''} (P31), participants think feature attribution \textbf{can answer how and why AI makes decisions}. Users also tend to assume features are independent of each other, and have causal effect towards the outcome.

For the \textit{attribution} information, users have various perceptions and requirements on the \textbf{level of detail}, including suggestions to set a user-defined threshold to \textit{``reduce the cognitive load''} (P04, \ul{learning}); and show feature ranking only, with the numeric attributions showing on demand.

Regarding applicable explanation goals, users tend to \textit{``\textbf{compare (AI's feature attribution) with my own judgment}, to see if that aligns with my feature ranking''} (P01, \ul{safety}). Users also use the feature importance
ranking 
to prioritize their action to \ul{improve user's predicted outcome}.

\subsubsection*{(2) Feature shape}
The graphical representation of showing  the relationship between a feature and the prediction makes users \textit{``feel so easy to latch onto, it's something that you can impact and something that’s very tangible''} (P22). Thus, many participants use it for \textbf{counterfactual reasoning} to \ul{improve user's outcome}. 
By showing the relationship between a protected feature (such as gender, ethnicity) and outcome,
feature shape is also helpful to \ul{reveal potential biases}.

For users' requirements, despite being a global explanation, participants expect the feature shape plot to \textbf{indicate the position of their input}. Furthermore, P30 suggested constructing
a \textbf{personalized feature shape explanation} by matching values of the remaining features (the feature set of all features except the targeted feature $t_i$ in the feature shape) with values from user's input features: \textit{``The AI should assume all other features are more similar to mine, considering this hypothesis then this is the (feature shape) curve''}. Participants also suggested attaching the feature shape explanation to other explanation forms as a detail and only showing it on demand.

\subsubsection*{(3) Feature interaction}
Despite being an extension of feature shape, four participants find the feature interaction graph \textit{``\textbf{less accessible to understand}''} (P22).
Similar to feature shape, feature interaction also supports counterfactual reasoning.
For users' requirements, participants would like to be able to select 
feature pairs to check their interactions, and expect AI to suggest or \textbf{prioritize feature pairs with significant or interesting interactions}.

\subsection{Example-based explanation}
\subsubsection*{(4) Similar example}
Similar example explanation \textit{``\textbf{intuitively makes sense}''} (P16) to participants,
as \textit{``it's similar to how humans make decisions''} (P02) by using analogical reasoning. 
\textit{``Even though these (similar and typical example) aren't much specific about how it (AI) is actually doing the (decision) process''} (P16), users, in their interpretation, automatically make up the model's reasoning process by comparing instances: \textit{``it gives you comparables, like the houses that are similar to yours, you can compare the price''} (P04).

To facilitate users' comparison among instances, users request the example-based explanation to \textbf{describe feature-level information}, \textit{``to pinpoint things (features) that are similar or different between these cases''} (P31), \textit{``I want to see if there are overlapped features''} (P01), \textit{``I need to know what factors AI is taking into account in order to compare it with the others''} (P16). In addition, the \textbf{feature representation should support users' side-by-side comparison} among multiple instances. This is especially vital when the input data format is high-dimensional or difficult to read through.

\subsubsection*{(5) Typical example}\label{te}
Participants may \textbf{not well distinguish the meaning} of a typical example from a similar example, and only seven participants explicitly understand the meaning of a typical example: \textit{``you’re getting the average''} (P20). For participants who understand its meaning, typical examples help them \ul{learn from AI} by revealing class-specific characteristics: \textit{``if you clearly separate each category, that helps people to differentiate the different categories''} (P04). It also helps to reveal potential problems in the AI model or data for the explanation goals such as to \ul{reveal bias}.

In some cases, 
participants \textbf{expect examples to convey a richer context}, such as several typical examples to represent within-class variations, or even \textbf{atypical or edge cases} that, albeit rare, may have stark consequences.
This is to understand AI capabilities due to \ul{safety} and \ul{biases} concerns: 
\textit{``So they (similar and typical example) don't really provide enough information about when the weather is different and when you’re driving at night, the results from non-typical conditions (do)''} (P27, Car task, \ul{Bias}).

\subsubsection*{(6) Counterfactual example}\label{ce}
It can serve different explanation goals depending on the context. In predictive tasks (House and Health), participants regard it as the straightforward explanation to suggest ways to \ul{improve users' predicted outcome}: \textit{``this one (counterfactual example) is the direct one telling you what to do to decrease your risk''} (P16, Health).
In recognition task (Bird), counterfactual example is suitable to show the differences to help users to \ul{learn} and \ul{differentiate} two similar predictions.

Two participants did \textbf{not understand the meaning} of counterfactual example, and could not capture the nuance between it and feature attribution, since they both have features highlighted but for different reasons. It may also make users confused about similar instances, especially in recognition tasks. 

Regarding users' requirements, the two contrastive \textit{predictions} in a counterfactual example can be \textbf{user-defined or pre-generated}, depending on the specific explanation goals. One prediction is usually from user’s input, and the alternative prediction could be: \textit{``the next possible prediction''} (P18), users’ own prediction when there is a \ul{decision disagreement}, the prospective prediction to \ul{improve users' outcome}, and the easily-confused prediction to \ul{differentiate similar instances}. 
In addition, the \textbf{\textit{counterfactual features} may also receive user-defined or pre-defined constraints}, such as  constraints on only including user-controllable features; and constraints on the range of features.

\subsection{Rule-based explanation}
\subsubsection*{(7) Decision rule}\label{dr}
Participants regard decision rule as able to \textit{``\textbf{explain the logic behind how the AI makes decisions}''} (P27). Particularly, the text description format is \textit{``like human explanation''} (P01), and \textit{``simple enough and understandable''} (P11). The text format can be regarded as \textit{``more exact''} (P05), and is helpful in \ul{report generation} and \ul{user learning}.

To reduce the cognitive load of complex rules and carefully balance between explanation completeness and usability, a few participants suggested \textbf{trimming the rules} by presenting shallow levels by default with a wide range of predictions, and showing more features and deep levels on demand with a narrow range of predictions; or \textit{``just show rules related to my own house features''} (P30); or highlight local rule clauses related to user's input on top of the global rule explanation.

\subsubsection*{(8) Decision tree}\label{dt}
Seven participants find it \textbf{difficult to interpret} the decision tree explanation. Four notice decision tree and decision rule provide  \textit{``basically the same information''} (P02), \textit{``all show the decision process''} (P10), and are only different in their text or graphical representation. Participants mention an advantage of decision tree is to \ul{differentiate similar instances} and \textbf{support counterfactual reasoning} by checking alternative feature values on adjacent branches. 
Similar to decision rule, participants tended to \textbf{focus on the branch pathway where their own input resides}.

\subsection{Contextual information}
\subsubsection*{(9) Input}
Participants regard input as a \textit{``profile''} (P24) that \textit{``stating the facts''} (P20). It allows participants to understand what information AI's decision is based on, and can help users to ``debug'' \textit{``if AI is missing the most important feature''}(P22), and \textit{``whether or not the input is enough for it (AI) to make that decision''} (P16). 

\subsubsection*{(10) Output} 
Participants have divergent interpretation on the presentation forms of AI's output prediction. Some prefer to see a prediction range because it \textit{``gives choices''}  (P05), \textit{``acknowledges a possibility''} (P18), ranks the priority of decisions (P03), helps them \textit{``to see how different between my and AI prediction''} (P01), and provides rooms for adjustment and negotiation. For some participants, seeing a point prediction or prediction with a narrower range may give them more confidence about AI's prediction.
We also include prediction uncertainty or confidence information in the output. Although a high certainty \textit{``reassure AI’s performance''} (P22), some participants find it difficult to interpret the level of uncertainty by its number without a range of reference. 

\subsubsection*{(11) Performance} 
After checking AI's performance, most participants realize the probabilistic nature of AI decisions: \textit{``AI is not perfect''} (P20), \textit{``they (AI) make errors sometimes''} (P05). 
Knowing the performance level helps users to set a proper expectation of AI's capabilities. In some particular explanation goals such as to \ul{detect bias}, participants may require to check the fine-grained performance analysis on subgroups.

\subsubsection*{(12) Dataset} 
Some participants tend to link the dataset size with model performance and their trust in AI:
\textit{``The higher the (training data distribution) curve goes, then I would be more confident that they have a big pool of data to pull from''}(P31).

\section{RQ2: End-users' requirements for explanation goals}

\subsection{Calibrating trust}\label{trust}
The process of user calibrating their trust in AI's current prediction and the model in general involves multiple factors and their complicated interactions. We summarize the following key emerged themes participants requested to calibrate their trust in AI.

    1. \textbf{Performance}:
Since end-users usually do not have complete computational and domain knowledge to judge AI’s decision process, model performance becomes an important surrogate to establish trust.
Prior work identified two types of performance: stated and observed performance~\cite{Yin2019}, and both were mentioned in the interview.
\textit{Stated performance} or accuracy is performance metrics tested on previous hold-out test data, and it was mentioned by most participants as a requirement to build trust in AI.
Compared to assessing the performance metrics, some participants tended to test AI by themselves and get hands-on \textit{observed performance} to be convinced. This requires users to have a referred ground truth from their own judgment or reliable external sources.

2. \textbf{Feature}: The important features that AI was based on are the next frequently mentioned information.
\textit{``I would like to know the list of criteria that the AI chose the price based on, and which one weighs more.''} (P30, House)

3. \textbf{The ability to discriminate similar instances}: This information was requested by several participants to demonstrate AI's capability.
\textit{``\textcolor{orange}{Typical example} seems to be pretty good at picking up on differences. \textcolor{orange}{Similar example} I can see that it's got a good variety of similar birds. So I found these ones make me trust it more.} (P16, Bird)

4. \textbf{Dataset}: The dataset size that AI was trained on is another surrogate mentioned by some participants to enhance trust. 
\textit{``If I know that the AI comes from a large database, it seems like the database is actually the experience that AI has. So the larger it (dataset) gets, the more experienced AI would be, so I can trust it more.''} (P30, House)

5. \textbf{External information}: 
This is 
another surrogate mentioned by participants to judge if AI is trustworthy. The external information could include: peer reviews, endorsement, and AI company's credit; and authority approval and liability.

\subsection{Ensuring safety}\label{safety}
To ensure the safety and reliability of the AI system in critical tasks (the autonomous driving vehicle task in our study), participants frequently mentioned \textbf{checking AI’s performances in test cases}, expecting the testing to \textbf{cover a variety of scenarios} to show the robustness of safety.
Although it is impossible to enumerate a complete list of potential failure cases in testing, extreme cases or potential accidents were the main concerns and focus of end-users.

Similar to the explanation goal to \textcolor{blue}{calibrate trust}, alongside the above \textit{stated performance}, a few participants required \textit{observed performance} to emotionally accept AI as an emerging technology.
\textit{``Definitely I would want to be in one car. I think information is not helpful, it’s not an intellectual factual thing, it's emotionally not acceptable. It (AI) is new and I have to learn to trust it.''} (P17)

\subsection{Detecting bias}\label{bias}
Participants were concerned about population bias~\cite{Mehrabi2019}, or distribution shift where AI models are applied to a different population other than the training dataset. Such concern is more prominent when a prediction is made on users’ own personal data, and when users are in minority subgroups. Participants wanted to compare and see \textbf{if their own subgroup is included in the training data}.

In cases where the AI task is not related to personal information (in our study the self-driving car task), participants required AI to \textbf{perform equally in all potentially biased conditions}.

\subsection{Unexpected prediction: when users disagree with AI}\label{unexpected}
When AI’s predictions did not align with participants’ own expectations, most participants would \textit{``question AI''} (P16, P20) and \textit{``the contradiction may let me confuse''} (P02). Some may lose trust in AI, thus would not go further to check its explanations, if they were confident about their own judgment.
Some would check \textit{``a trusted second opinion''} (P06, P10), or refer to human experts (P12).

But for the majority of participants, \textbf{explanations were needed \textit{``to know why''}} (P01) and to resolve conflicts.
Explanations
help users to identify AI’s flaws and reject AI, or to be convinced and adjust user's own judgment by checking detailed differences on rationale, although \textit{``it might be harder to persuade me''} (P31). Specifically, participants  
\textit{``try to understand what makes a difference (between AI and my prediction)''} (P03), which is similar to the explanation goal to \textcolor{blue}{differentiate} (Section~\ref{differentiate}). 
To show why the predictions are different, many participants \textbf{required a list of key features}.
In the case that AI made errors, seeing what AI is based on can facilitate user's ``debugging’’ process. 
Although end-users cannot debug the algorithmic part, they may be able to debug the input to see if AI \textit{``have the complete information''} (P03) as users have.
Furthermore, if some key input information is lacking in AI’s decision, the system needs to allow users to provide feedback by inputting more information (P03, P24), or \textit{``correct the error''} (P16) for AI.

\subsection{Expected prediction: when users agree with AI} 
When the prediction matched participants’ expectations, participants \textit{``will trust the AI more''} (P10), and the motivation to check explanations was \textit{``not as strong as the previous one (\textcolor{blue}{unexpected})''} (P02). Some participants stopped at the prediction, willing to accept the ``black-box’’ AI and may \textit{``not even waste my time (checking explanations)''} (P20). 
A few participants still wanted to check further explanations to boost user's confidence, or improve user's predicted outcome.

\subsection{Differentiating similar instances}\label{differentiate}
To facilitate end-users’ explanation goal to differentiate similar instances, an AI system is required to first have the ability to discern similar instances.
If AI is doubtful about its predictions for similar instances, participants expected AI to indicate its prediction uncertainty.
Built upon the above factors, AI needs to be able to \textit{``\textbf{pinpoint unique features} that made them really different from each other''} (P06). And the user interface may additionally support users’ own side-by-side comparison for the similar instances.

\subsection{Learning from AI}\label{learn}
Using AI for user's personal learning, improving problem-solving skills, and knowledge discovery \textit{``depends on how reliable it (AI) really is''} (P10).
To facilitate learning and knowledge discovery, 
participants \textbf{expected a wide range of explanations} depending on the particular learning goal, such as \textit{``more details to systematically learn, go over that same bird, ...a mind map to build a category of birds by one feature''} (P02), \textit{``the specific characteristic about this bird, and how can I differentiate this bird from other birds''} (P04).

\subsection{Improving user's predicted outcome}\label{control}

Participants intuitively sought explanations to improve their predicted outcome when predictions are related to personal data (in our study, the House and Health tasks).
Regarding the requested explanatory information,
participants were looking for suggestions with \textbf{controllable features} and
ignored explanations that include uncontrollable features.
Knowing the controllable features has a positive psychological effect to give users a sense of control, and vice versa.

\subsection{Communicating with stakeholders}\label{communicate}
To communicate with other stakeholders, some participants chose to communicate verbally about their opinions without mentioning AI. Others preferred to present stakeholders with more evidence by bringing AI's additional information explicitly to the discussion. For the latter case, the \textbf{other stakeholders need to establish basic understanding and trust in AI} before discussing AI’s explanation.
To do so, most participants chose to present AI’s \textcolor{orange}{performance} information first to build trust.
A \textbf{formal summary or report from AI} may facilitate the communication with other stakeholders, as requested by many participants.

\subsection{Generating reports}\label{report}
The content of reports may largely depend on the specific explanation goal and targeted readers of the report. In our study, 
participants frequently mentioned the report should include\textit{``key identifying features''},\textit{ ``list of distinguishing characteristics or what makes it unique''} (P09),
or \textit{``\textbf{a summary of factors} that were part of the input led to the diabetic prediction''} (P31). Users also mentioned including contextual information to back up the decisions, such as the training dataset size of the predicted class, and the level of decision certainty (P01).

\subsection{Multiple objectives trade-off}\label{multi}

Usually it is the human user rather than AI to trade off among multiple objectives in AI-assisted decision-making tasks. Thus, when multiple objectives conflict (in our study, they are the scenarios when car drives autonomously and passenger gets a car sick; AI predicts diabetes and uses it to determine insurance premium), AI is required to allow users to take over or to receive users’ inputs.
Explanations are required if the multiple objectives conflict and need to trade off. 
And users could use such explanations to defend for or against certain objectives.

\section{Discussion}
We discuss three aspects to illustrate how the EUCA study results on explanation forms and goals can be applied to support the design and evaluation of end-user-centered XAI techniques.

\subsection{Designing new XAI techniques based on users' requirements for explanation forms}\label{form_design}
Current research on XAI algorithms are technical-user-centered, which disproportionately ignores non-technical end-users' demand for XAI. To bridge the research gap, a recent work suggests the end-user-inspired design to model new XAI technical problems using end-users' requirements from EUCA as a source of inspiration~\cite{transcending}. It proposes three novel XAI algorithms that address various end-users' explanation goals: 
\textbf{1}. The egde-case-based reasoning algorithm is a new form of example-based explanation, proposed for the explanation goals to ensure safety (Section~\ref{safety}). It can identify edge cases that are wrongly predicted by AI and pose significant consequences. It was inspired by the EUCA study finding on typical examples, where participants requested to check untypical or edge cases for safety concerns
(Section~\ref{te}).
\textbf{2}. The collapsible decision tree algorithm directly addressed users' requirements for rule-based explanation (Section~\ref{dr}) to balance the conciseness and comprehensiveness of a decision tree and improve its usability.
\textbf{3}. The user-customizable counterfactual explanation algorithm enables users to customize a counterfactual example by setting changeable and unchangeable features and their ranges. The algorithm was motivated by users' requirements on controllable features (Section~\ref{ce}) for their explanation goal to improve user's predicted outcome (Section~\ref{control}). The authors also noted that the design of this XAI technique was closely tied to its top required explanation goals in Fig.~\ref{fig:mtx}, because the EUCA study and prior user study~\cite{Cai2019} have found users' may get confused with counterfactual examples when using it for other explanation goals, such as model understanding. 

From these examples of EUCA-inspired new XAI techniques, we can observe that because of the bridging role of explanation forms that connect users' requirements to the backend XAI techniques, the users' requirements for explanation forms can be used to directly formulate new XAI techniques.
In addition, users' requirements from the EUCA study are acquired in the context of explanation goals. This makes the design of new XAI techniques more targeted for users' intended purposes. 

\subsection{Optimizing XAI techniques for explanation form}
From the abstraction of XAI techniques by its explanation form and users' selection of explanation forms, we suggest that XAI techniques can be evaluated and optimized not only by its explanation content, but also by its explanation form. 
For example, the clinical XAI guidelines recommend using the end-user-friendly explanation forms from EUCA to assess and select the optimal explanation form or the combination of forms with clinical users, so that to meet the understandability and clinical relevance criteria in the guidelines~\cite{JIN2023102684}. Optimizing XAI by its explanation form can  make the explanation more complete with complementary reasoning process from different forms. Users' opinions are also involved in the optimization process to make the XAI techniques more understandable and user-friendly. 
In addition to the above way of optimization by selection and combination of explanation forms, as a direct future work of the EUCA study, Jin et al. suggested another way to optimize explanation form~\cite{transcending}. They dissected an explanation form as two components of human-interpretable features and a reasoning process, and the form of representing human-interpretable features can also be optimized.
More future work can explore how to optimize explanation form and its feature presentation for more comprehensive~\cite{rethinking} or even interactive and personalized explanation~\cite{Schneider2019PersonalizedEI}. We make the EUCA quantitative data available ($N=248$), including each participant's information and their selections of explanation forms for each explanation goal, to support the modeling of personalized explanation.

\subsection{Evaluating XAI techniques based on users' requirements for explanation goals}\label{goal_eval}

New user-centered XAI techniques should be evaluated regarding human-grounded properties~\cite{10.1145/3387166} that end-users are concerned about. However, it is difficult to decide what to evaluate and how to evaluate it when designing end-user-oriented XAI. 
Regarding what to evaluate, we suggest that the focus of both computational and human-subject evaluations should be on the utility of XAI techniques to end-users~\cite{rethinking}, including the direct XAI utility to users' explanation goals, such as how effectively the XAI techniques can help users to resolve their disagreement, calibrate trust to take or reject AI's decision~\cite{NEURIPS2021_de043a5e, shen2020useful,Kim2022HIVE}, or detect biases~\cite{adebayo2022post}; and the indirect utility of XAI to user's entire task, such as human-AI team task performance~\cite{Jin2022.12.07.22282726, taesiri2022visual, siu2021evaluation}.
Regarding how to evaluate, the EUCA finding on users' requirements for explanation goals provides dynamics on user's information synthesis process, which can inspire the design of human-subject evaluation and the formulation of computational metrics. For example, inspired by the EUCA study findings on explanation goals, Jin et al. proposed an evaluation metric verifiability to computationally evaluate the XAI utility of verifying AI decisions~\cite{transcending}. They referred to the identified process in EUCA that users calibrate their trust to take or reject AI decisions by comparing important features from AI explanation with their prior knowledge (Section~\ref{trust}), and abstracted such process by the metric verifiability. The computational metric acts as a surrogate and complement of human-subject evaluation, and it improves the efficiency of XAI evaluation and can be used to directly optimize XAI algorithms. With the understanding of users' requirements for explanation goals, future work can develop more computational or human-subject evaluations to improve the utility of XAI techniques on various end-user-oriented explanation goals.

\section{Limitation and future work}
Our study discovers some typical responses and requirements that end-users would demand for XAI, but it is not a complete list of user requirements. Limited by the scope of a single study, we have to trade off the depth of exploring a single form or goal, for the breadth of uncovering user requirements for a number of common explanation forms and goals in the context of their comparisons and interactions.
We hope this work can inspire more future works that construct a closed loop in the research and development of user-centered XAI techniques:

\textbf{1}. User understandings improve XAI techniques: we hope the user requirements identified in our and other studies can directly inspire the research and development of end-user-centered XAI techniques. We encourage more bridge works such as~\cite{transcending} that bridge the AI community with HCI advances to inform XAI researchers and designers of end-users' requirements, and propose XAI techniques and evaluation methods driven by end-users' requirements. Our study provides useful tools to focus user requirements on explanation forms and goals to effectively convert their requirements into technical specifications.

\textbf{2}. XAI techniques motivate more research questions on user understanding: to promote the research and development of improving a particular XAI algorithm or interface (manifested by its explanation form), future research needs to conduct more user studies to understand the in-depth domain- or task-specific user requirements for a particular explanation form and its interaction with one or several explanation goals. To promote research on improving an XAI evaluation method (manifested by its explanation goal), future work needs to conduct more targeted user studies to quantify and model the relationship between the quantifiable factors and the explanation goal, whereas in our work we only qualitatively identify certain factors, such as the five factors to the explanation goal of calibrating trust (Section~\ref{trust}). The emerging questions from the technical development of user-centered XAI can motivate new user studies, and their findings will in turn promote the technical development of user-centered XAI.

\section{Conclusion}
Understanding end-users' requirements is an essential but challenging task to develop XAI techniques that are useful for end-users. To effectively acquire users requirements to support the development of XAI techniques, we propose the end-user-friendly explanation forms as a communication interface between XAI techniques and end-users. We also identify the common explanation goals that motivate an end-users to seek explanation. With these, we conduct the EUCA user study with 32 layperson participants in four AI-assisted critical tasks. The results identify end-users' requirements for explanation forms and goals in the context of their interactions and comparisons. We show with examples that the user study results on user requirements for explanation forms can inspire the design of XAI techniques, and the findings on user requirements for explanation goals can inform the evaluation of XAI techniques. The EUCA user study findings, the identified explanation forms and goals for technical specification, and the EUCA study dataset support the incorporation of end-users' requirements in the design and evaluation of end-user-centered XAI techniques, which aim to make AI accessible, safe, and accountable to use for end-users.

\section*{Research Ethics \& Social Impact}
The study was approved by the university's Research Ethics Board (Ethics number: 2019s0244). No personal identifiable information was associated with the data. We make our study materials and user study data publicly available to enable the reproducibility and transparency of this work.
The user study design aims to make AI more accessible and equal to non-technical end-users. With a particular focus on end-users' demand for XAI in their critical tasks, the study also supports the design of safe and responsible AI in high-stakes applications. 

\section*{Acknowledgements}
We thank all study participants for their time, effort, and valuable input in the study. We thank Sheelagh Carpendale, Parmit Chilana, Ben Cardoen, Pegah Kiaei, Zipeng Liu, and Mingbo Cai for the helpful discussions in shaping this work. The work was supported by Simon Fraser University Big Data Initiative The Next Big Question Funding. 

\bibliographystyle{icml2022}
\bibliography{xai, xai_old}

\begin{thebibliography}{81}
\providecommand{\natexlab}[1]{#1}
\providecommand{\url}[1]{\texttt{#1}}
\expandafter\ifx\csname urlstyle\endcsname\relax
  \providecommand{\doi}[1]{doi: #1}\else
  \providecommand{\doi}{doi: \begingroup \urlstyle{rm}\Url}\fi

\bibitem[gdp()]{gdpr}
{The impact of the General Data Protection Regulation (GDPR) on artificial
  intelligence}.
\newblock \doi{10.2861/293}.
\newblock URL \url{http://www.europarl.europa.eu/thinktank}.

\bibitem[Adebayo et~al.(2022)Adebayo, Muelly, Abelson, and
  Kim]{adebayo2022post}
Adebayo, J., Muelly, M., Abelson, H., and Kim, B.
\newblock Post hoc explanations may be ineffective for detecting unknown
  spurious correlation.
\newblock In \emph{International Conference on Learning Representations}, 2022.
\newblock URL \url{https://openreview.net/forum?id=xNOVfCCvDpM}.

\bibitem[Alicioglu \& Sun(2022)Alicioglu and Sun]{ALICIOGLU2022502}
Alicioglu, G. and Sun, B.
\newblock A survey of visual analytics for explainable artificial intelligence
  methods.
\newblock \emph{Computers \& Graphics}, 102:\penalty0 502--520, 2022.
\newblock ISSN 0097-8493.
\newblock \doi{https://doi.org/10.1016/j.cag.2021.09.002}.
\newblock URL
  \url{https://www.sciencedirect.com/science/article/pii/S0097849321001886}.

\bibitem[Apley(2016)]{Apley2016}
Apley, D.~W.
\newblock {Visualizing the Effects of Predictor Variables in Black Box
  Supervised Learning Models}.
\newblock dec 2016.
\newblock URL \url{http://arxiv.org/abs/1612.08468}.

\bibitem[Bhatt et~al.(2020)Bhatt, Xiang, Sharma, Weller, Taly, Jia, Ghosh,
  Puri, Moura, and Eckersley]{Bhatt2019}
Bhatt, U., Xiang, A., Sharma, S., Weller, A., Taly, A., Jia, Y., Ghosh, J.,
  Puri, R., Moura, J.~M., and Eckersley, P.
\newblock {Explainable machine learning in deployment}.
\newblock In \emph{FAT* 2020 - Proceedings of the 2020 Conference on Fairness,
  Accountability, and Transparency}, pp.\  648--657, 2020.
\newblock ISBN 9781450369367.
\newblock \doi{10.1145/3351095.3375624}.
\newblock URL \url{http://arxiv.org/abs/1909.06342}.

\bibitem[Braun \& Clarke(2012)Braun and Clarke]{Braun2012}
Braun, V. and Clarke, V.
\newblock {Thematic analysis.}
\newblock In \emph{APA handbook of research methods in psychology, Vol 2:
  Research designs: Quantitative, qualitative, neuropsychological, and
  biological.}, APA handbooks in psychology{\textregistered}., pp.\  57--71.
  American Psychological Association, Washington, DC, US, 2012.
\newblock ISBN 1-4338-1005-0 (Hardcover); 978-1-43381-005-3 (Hardcover).
\newblock \doi{10.1037/13620-004}.

\bibitem[Bussone et~al.(2015)Bussone, Stumpf, and O'Sullivan]{7349687}
Bussone, A., Stumpf, S., and O'Sullivan, D.
\newblock The role of explanations on trust and reliance in clinical decision
  support systems.
\newblock In \emph{2015 International Conference on Healthcare Informatics},
  pp.\  160--169, 2015.
\newblock \doi{10.1109/ICHI.2015.26}.

\bibitem[Cai et~al.(2019{\natexlab{a}})Cai, Jongejan, and Holbrook]{Cai2019}
Cai, C.~J., Jongejan, J., and Holbrook, J.
\newblock {The effects of example-based explanations in a machine learning
  interface}.
\newblock In \emph{Proceedings of the 24th International Conference on
  Intelligent User Interfaces - IUI '19}, pp.\  258--262, New York, New York,
  USA, 2019{\natexlab{a}}. ACM Press.
\newblock ISBN 9781450362726.
\newblock \doi{10.1145/3301275.3302289}.
\newblock URL \url{http://dl.acm.org/citation.cfm?doid=3301275.3302289}.

\bibitem[Cai et~al.(2019{\natexlab{b}})Cai, Reif, Hegde, Hipp, Kim, Smilkov,
  Wattenberg, Viegas, Corrado, Stumpe, and Terry]{10.1145/3290605.3300234}
Cai, C.~J., Reif, E., Hegde, N., Hipp, J., Kim, B., Smilkov, D., Wattenberg,
  M., Viegas, F., Corrado, G.~S., Stumpe, M.~C., and Terry, M.
\newblock Human-centered tools for coping with imperfect algorithms during
  medical decision-making.
\newblock In \emph{Proceedings of the 2019 CHI Conference on Human Factors in
  Computing Systems}, CHI '19, pp.\  1–14, New York, NY, USA,
  2019{\natexlab{b}}. Association for Computing Machinery.
\newblock \doi{10.1145/3290605.3300234}.
\newblock URL \url{https://doi.org/10.1145/3290605.3300234}.

\bibitem[Caruana et~al.(2015)Caruana, Lou, Gehrke, Koch, Sturm, and
  Elhadad]{Caruana2015}
Caruana, R., Lou, Y., Gehrke, J., Koch, P., Sturm, M., and Elhadad, N.
\newblock {Intelligible models for healthcare: Predicting pneumonia risk and
  hospital 30-day readmission}.
\newblock In \emph{Proceedings of the ACM SIGKDD International Conference on
  Knowledge Discovery and Data Mining}, volume 2015-Augus, pp.\  1721--1730,
  New York, New York, USA, aug 2015. Association for Computing Machinery.
\newblock ISBN 9781450336642.
\newblock \doi{10.1145/2783258.2788613}.
\newblock URL \url{http://dl.acm.org/citation.cfm?doid=2783258.2788613}.

\bibitem[Chazette et~al.(2022)Chazette, Kl\"{u}nder, Balci, and
  Schneider]{Chazette2022}
Chazette, L., Kl\"{u}nder, J., Balci, M., and Schneider, K.
\newblock How can we develop explainable systems? insights from a literature
  review and an interview study.
\newblock In \emph{Proceedings of the International Conference on Software and
  System Processes and International Conference on Global Software
  Engineering}. {ACM}, May 2022.
\newblock \doi{10.1145/3529320.3529321}.
\newblock URL \url{https://doi.org/10.1145/3529320.3529321}.

\bibitem[Chen et~al.(2019)Chen, Li, Tao, Barnett, Rudin, and Su]{Chen}
Chen, C., Li, O., Tao, D., Barnett, A., Rudin, C., and Su, J.~K.
\newblock This looks like that: Deep learning for interpretable image
  recognition.
\newblock In Wallach, H., Larochelle, H., Beygelzimer, A., d\textquotesingle
  Alch\'{e}-Buc, F., Fox, E., and Garnett, R. (eds.), \emph{Advances in Neural
  Information Processing Systems}, volume~32. Curran Associates, Inc., 2019.
\newblock URL
  \url{https://proceedings.neurips.cc/paper/2019/file/adf7ee2dcf142b0e11888e72b43fcb75-Paper.pdf}.

\bibitem[Cirqueira et~al.(2020)Cirqueira, Nedbal, Helfert, and
  Bezbradica]{fraud}
Cirqueira, D., Nedbal, D., Helfert, M., and Bezbradica, M.
\newblock Scenario-based requirements elicitation for user-centric explainable
  ai.
\newblock In Holzinger, A., Kieseberg, P., Tjoa, A.~M., and Weippl, E. (eds.),
  \emph{Machine Learning and Knowledge Extraction}, pp.\  321--341, Cham, 2020.
  Springer International Publishing.
\newblock ISBN 978-3-030-57321-8.

\bibitem[Dhurandhar et~al.(2018)Dhurandhar, Chen, Luss, Tu, Ting, Shanmugam,
  and Das]{Dhurandhar}
Dhurandhar, A., Chen, P.-Y., Luss, R., Tu, C.-C., Ting, P., Shanmugam, K., and
  Das, P.
\newblock Explanations based on the missing: Towards contrastive explanations
  with pertinent negatives.
\newblock In \emph{Proceedings of the 32nd International Conference on Neural
  Information Processing Systems}, NIPS'18, pp.\  590–601, Red Hook, NY, USA,
  2018. Curran Associates Inc.

\bibitem[Doshi-Velez \& Kim(2017)Doshi-Velez and Kim]{Doshi-Velez2017}
Doshi-Velez, F. and Kim, B.
\newblock {Towards A Rigorous Science of Interpretable Machine Learning}.
\newblock feb 2017.
\newblock URL \url{http://arxiv.org/abs/1702.08608}.

\bibitem[Evans et~al.(2022)Evans, Retzlaff, Geißler, Kargl, Plass, Müller,
  Kiehl, Zerbe, and Holzinger]{EVANS2022281}
Evans, T., Retzlaff, C.~O., Geißler, C., Kargl, M., Plass, M., Müller, H.,
  Kiehl, T.-R., Zerbe, N., and Holzinger, A.
\newblock The explainability paradox: Challenges for xai in digital pathology.
\newblock \emph{Future Generation Computer Systems}, 133:\penalty0 281--296,
  2022.
\newblock ISSN 0167-739X.
\newblock \doi{https://doi.org/10.1016/j.future.2022.03.009}.
\newblock URL
  \url{https://www.sciencedirect.com/science/article/pii/S0167739X22000838}.

\bibitem[Ferreira \& Monteiro(2020)Ferreira and Monteiro]{Ferreira2020}
Ferreira, J.~J. and Monteiro, M.~S.
\newblock {What Are People Doing About XAI User Experience? A Survey on AI
  Explainability Research and Practice}.
\newblock pp.\  56--73. Springer, Cham, jul 2020.
\newblock \doi{10.1007/978-3-030-49760-6_4}.
\newblock URL \url{http://link.springer.com/10.1007/978-3-030-49760-6{\_}4}.

\bibitem[Friedman(2001)]{Friedman2001}
Friedman, J.~H.
\newblock {Greedy function approximation: A gradient boosting machine.}
\newblock \emph{The Annals of Statistics}, 29\penalty0 (5):\penalty0
  1189--1232, oct 2001.
\newblock \doi{10.1214/aos/1013203451}.
\newblock URL \url{http://projecteuclid.org/euclid.aos/1013203451}.

\bibitem[Frosst \& Hinton(2017)Frosst and Hinton]{Frosst}
Frosst, N. and Hinton, G.
\newblock {Distilling a Neural Network Into a Soft Decision Tree}.
\newblock Technical report, 2017.
\newblock URL \url{https://arxiv.org/pdf/1711.09784.pdf}.

\bibitem[Goyal et~al.(2019)Goyal, Wu, Ernst, Batra, Parikh, and
  Lee]{pmlr-v97-goyal19a}
Goyal, Y., Wu, Z., Ernst, J., Batra, D., Parikh, D., and Lee, S.
\newblock Counterfactual visual explanations.
\newblock In Chaudhuri, K. and Salakhutdinov, R. (eds.), \emph{Proceedings of
  the 36th International Conference on Machine Learning}, volume~97 of
  \emph{Proceedings of Machine Learning Research}, pp.\  2376--2384. PMLR,
  09--15 Jun 2019.
\newblock URL \url{https://proceedings.mlr.press/v97/goyal19a.html}.

\bibitem[Gregor \& Benbasat(1999)Gregor and Benbasat]{Gregor1999a}
Gregor, S. and Benbasat, I.
\newblock {Explanations from Intelligent Systems: Theoretical Foundations and
  Implications for Practice}.
\newblock \emph{MIS Quarterly}, 23\penalty0 (4):\penalty0 497, dec 1999.
\newblock ISSN 02767783.
\newblock \doi{10.2307/249487}.
\newblock URL \url{https://www.jstor.org/stable/249487?origin=crossref}.

\bibitem[Guidotti et~al.(2018)Guidotti, Monreale, Ruggieri, Pedreschi, Turini,
  and Giannotti]{Guidotti2018c}
Guidotti, R., Monreale, A., Ruggieri, S., Pedreschi, D., Turini, F., and
  Giannotti, F.
\newblock {Local Rule-Based Explanations of Black Box Decision Systems}.
\newblock may 2018.
\newblock URL \url{http://arxiv.org/abs/1805.10820}.

\bibitem[Habiba et~al.(2022)Habiba, Bogner, and Wagner]{9920137}
Habiba, U.-E., Bogner, J., and Wagner, S.
\newblock Can requirements engineering support explainable artificial
  intelligence? towards a user-centric approach for explainability
  requirements.
\newblock In \emph{2022 IEEE 30th International Requirements Engineering
  Conference Workshops (REW)}, pp.\  162--165, 2022.
\newblock \doi{10.1109/REW56159.2022.00038}.

\bibitem[Holzinger et~al.(2017)Holzinger, Malle, Kieseberg, Roth, M{\"{u}}ller,
  Reihs, and Zatloukal]{Holzinger2017a}
Holzinger, A., Malle, B., Kieseberg, P., Roth, P.~M., M{\"{u}}ller, H., Reihs,
  R., and Zatloukal, K.
\newblock {Towards the Augmented Pathologist: Challenges of Explainable-AI in
  Digital Pathology}.
\newblock dec 2017.
\newblock URL \url{http://arxiv.org/abs/1712.06657}.

\bibitem[Hong et~al.(2020)Hong, Hullman, and Bertini]{Hong2020}
Hong, S.~R., Hullman, J., and Bertini, E.
\newblock {Human Factors in Model Interpretability: Industry Practices,
  Challenges, and Needs}.
\newblock \emph{Proceedings of the ACM on Human-Computer Interaction},
  4\penalty0 (CSCW1):\penalty0 1--26, 2020.
\newblock ISSN 25730142.
\newblock \doi{10.1145/3392878}.

\bibitem[Hooper \& Hsia(1982)Hooper and Hsia]{scenario}
Hooper, J.~W. and Hsia, P.
\newblock Scenario-based prototyping for requirements identification.
\newblock \emph{SIGSOFT Softw. Eng. Notes}, 7\penalty0 (5):\penalty0 88–93,
  April 1982.
\newblock ISSN 0163-5948.
\newblock \doi{10.1145/1006258.1006275}.
\newblock URL \url{https://doi.org/10.1145/1006258.1006275}.

\bibitem[Huysmans et~al.(2011)Huysmans, Dejaeger, Mues, Vanthienen, and
  Baesens]{10.1016/j.dss.2010.12.003}
Huysmans, J., Dejaeger, K., Mues, C., Vanthienen, J., and Baesens, B.
\newblock An empirical evaluation of the comprehensibility of decision table,
  tree and rule based predictive models.
\newblock \emph{Decis. Support Syst.}, 51\penalty0 (1):\penalty0 141–154, apr
  2011.
\newblock ISSN 0167-9236.
\newblock \doi{10.1016/j.dss.2010.12.003}.
\newblock URL \url{https://doi.org/10.1016/j.dss.2010.12.003}.

\bibitem[Jacobs et~al.(2021)Jacobs, Pradier, McCoy, Perlis, Doshi-Velez, and
  Gajos]{Jacobs2021}
Jacobs, M., Pradier, M.~F., McCoy, T.~H., Perlis, R.~H., Doshi-Velez, F., and
  Gajos, K.~Z.
\newblock How machine-learning recommendations influence clinician treatment
  selections: the example of antidepressant selection.
\newblock \emph{Translational Psychiatry}, 11\penalty0 (1), February 2021.
\newblock \doi{10.1038/s41398-021-01224-x}.
\newblock URL \url{https://doi.org/10.1038/s41398-021-01224-x}.

\bibitem[Jin et~al.(2019)Jin, Carpendale, Hamarneh, and Gromala]{Jin2019vis}
Jin, W., Carpendale, S., Hamarneh, G., and Gromala, D.
\newblock {Bridging AI Developers and End Users: an End-User-Centred
  Explainable AI Taxonomy and Visual Vocabularies}.
\newblock In \emph{IEEE VIS 2019 Conference Poster Abstract}, 2019.

\bibitem[Jin et~al.(2020)Jin, Fatehi, Abhishek, Mallya, Toyota, and
  Hamarneh]{Jin2020}
Jin, W., Fatehi, M., Abhishek, K., Mallya, M., Toyota, B., and Hamarneh, G.
\newblock {Artificial intelligence in glioma imaging: challenges and advances}.
\newblock \emph{Journal of neural engineering}, 17\penalty0 (2):\penalty0
  021002, 2020.
\newblock ISSN 17412552.
\newblock \doi{10.1088/1741-2552/ab8131}.

\bibitem[Jin et~al.(2022{\natexlab{a}})Jin, Fan, Gromala, Pasquier, Li, and
  Hamarneh]{transcending}
Jin, W., Fan, J., Gromala, D., Pasquier, P., Li, X., and Hamarneh, G.
\newblock Transcending {XAI} algorithm boundaries through end-user-inspired
  design, 2022{\natexlab{a}}.
\newblock URL \url{https://arxiv.org/abs/2208.08739}.

\bibitem[Jin et~al.(2022{\natexlab{b}})Jin, Fatehi, Guo, and
  Hamarneh]{Jin2022.12.07.22282726}
Jin, W., Fatehi, M., Guo, R., and Hamarneh, G.
\newblock Evaluating the clinical utility of artificial intelligence assistance
  and its explanation on glioma grading task.
\newblock \emph{medRxiv}, 2022{\natexlab{b}}.
\newblock \doi{10.1101/2022.12.07.22282726}.
\newblock URL
  \url{https://www.medrxiv.org/content/early/2022/12/09/2022.12.07.22282726}.

\bibitem[Jin et~al.(2022{\natexlab{c}})Jin, Li, and
  Hamarneh]{Jin_Li_Hamarneh_2022}
Jin, W., Li, X., and Hamarneh, G.
\newblock Evaluating explainable ai on a multi-modal medical imaging task: Can
  existing algorithms fulfill clinical requirements?
\newblock \emph{Proceedings of the AAAI Conference on Artificial Intelligence},
  36\penalty0 (11):\penalty0 11945--11953, Jun. 2022{\natexlab{c}}.
\newblock \doi{10.1609/aaai.v36i11.21452}.
\newblock URL \url{https://ojs.aaai.org/index.php/AAAI/article/view/21452}.

\bibitem[Jin et~al.(2023{\natexlab{a}})Jin, Li, Fatehi, and
  Hamarneh]{JIN2023102684}
Jin, W., Li, X., Fatehi, M., and Hamarneh, G.
\newblock Guidelines and evaluation of clinical explainable ai in medical image
  analysis.
\newblock \emph{Medical Image Analysis}, 84:\penalty0 102684,
  2023{\natexlab{a}}.
\newblock ISSN 1361-8415.
\newblock \doi{https://doi.org/10.1016/j.media.2022.102684}.
\newblock URL
  \url{https://www.sciencedirect.com/science/article/pii/S1361841522003127}.

\bibitem[Jin et~al.(2023{\natexlab{b}})Jin, Li, and Hamarneh]{rethinking}
Jin, W., Li, X., and Hamarneh, G.
\newblock Rethinking {AI} explainability and plausibility.
\newblock 2023{\natexlab{b}}.

\bibitem[Kawahara et~al.(2017)Kawahara, Moriarty, and Hamarneh]{Kawaharaa}
Kawahara, J., Moriarty, K.~P., and Hamarneh, G.
\newblock {Graph geodesics to find progressively similar skin lesion images}.
\newblock In \emph{Lecture Notes in Computer Science (including subseries
  Lecture Notes in Artificial Intelligence and Lecture Notes in
  Bioinformatics)}, volume 10551 LNCS, pp.\  31--41, 2017.
\newblock ISBN 9783319676746.
\newblock \doi{10.1007/978-3-319-67675-3_4}.
\newblock URL
  \url{https://www2.cs.sfu.ca/{~}hamarneh/ecopy/miccai{\_}grail2017.pdf}.

\bibitem[Kim et~al.(2016)Kim, Khanna, and Koyejo]{Kim2016}
Kim, B., Khanna, R., and Koyejo, O.~O.
\newblock Examples are not enough, learn to criticize! criticism for
  interpretability.
\newblock In Lee, D.~D., Sugiyama, M., Luxburg, U.~V., Guyon, I., and Garnett,
  R. (eds.), \emph{Advances in Neural Information Processing Systems 29}, pp.\
  2280--2288. Curran Associates, Inc., 2016.
\newblock URL
  \url{http://papers.nips.cc/paper/6300-examples-are-not-enough-learn-to-criticize-criticism-for-interpretability.pdf}.

\bibitem[Kim et~al.(2018)Kim, Wattenberg, Gilmer, Cai, Wexler, Viegas, and
  sayres]{Kim}
Kim, B., Wattenberg, M., Gilmer, J., Cai, C., Wexler, J., Viegas, F., and
  sayres, R.
\newblock Interpretability beyond feature attribution: Quantitative testing
  with concept activation vectors ({TCAV}).
\newblock volume~80 of \emph{Proceedings of Machine Learning Research}, pp.\
  2668--2677, Stockholmsmässan, Stockholm Sweden, 10--15 Jul 2018. PMLR.
\newblock URL \url{http://proceedings.mlr.press/v80/kim18d.html}.

\bibitem[Kim et~al.(2022)Kim, Meister, Ramaswamy, Fong, and
  Russakovsky]{Kim2022HIVE}
Kim, S. S.~Y., Meister, N., Ramaswamy, V.~V., Fong, R., and Russakovsky, O.
\newblock {HIVE}: Evaluating the human interpretability of visual explanations.
\newblock In \emph{European Conference on Computer Vision (ECCV)}, 2022.

\bibitem[Kleinberg et~al.(2017)Kleinberg, Lakkaraju, Leskovec, Ludwig,
  Mullainathan, Abrams, Alsdorf, Cohen, Crohn, Cusick, Dierks, Donohue, Dupont,
  Egan, Glazer, Gottschall, Hess, Kane, Kellam, LascalaGruenewald, Loeffler,
  Milgram, Raphael, Rohlfs, Rosenbaum, Salo, Shleifer, Sojourner, Sowerby,
  Sunstein, Sviridoff, Turner, and John]{Kleinberg2017}
Kleinberg, J., Lakkaraju, H., Leskovec, J., Ludwig, J., Mullainathan, S.,
  Abrams, D., Alsdorf, M., Cohen, M., Crohn, A., Cusick, G.~R., Dierks, T.,
  Donohue, J., Dupont, M., Egan, M., Glazer, E., Gottschall, J., Hess, N.,
  Kane, K., Kellam, L., LascalaGruenewald, A., Loeffler, C., Milgram, A.,
  Raphael, L., Rohlfs, C., Rosenbaum, D., Salo, T., Shleifer, A., Sojourner,
  A., Sowerby, J., Sunstein, C., Sviridoff, M., Turner, E., and John, J.
\newblock {Human Decisions and Machine Predictions}.
\newblock \penalty0 (September):\penalty0 1--53, 2017.
\newblock
  \doi{10.1093/qje/qjx032/4095198/Human-Decisions-and-Machine-Predictions}.
\newblock URL
  \url{https://academic.oup.com/qje/article-abstract/doi/10.1093/qje/qjx032/4095198/Human-Decisions-and-Machine-Predictions}.

\bibitem[Kolodner(1992)]{Kolodner1992}
Kolodner, J.~L.
\newblock {An introduction to case-based reasoning}.
\newblock \emph{Artificial Intelligence Review}, 6\penalty0 (1):\penalty0
  3--34, mar 1992.
\newblock ISSN 02692821.
\newblock \doi{10.1007/BF00155578}.
\newblock URL \url{https://link.springer.com/article/10.1007/BF00155578}.

\bibitem[Laato et~al.(2022)Laato, Tiainen, Islam, and
  M\"{a}ntym\"{a}ki]{Laato2022}
Laato, S., Tiainen, M., Islam, A.~N., and M\"{a}ntym\"{a}ki, M.
\newblock How to explain {AI} systems to end users: a systematic literature
  review and research agenda.
\newblock \emph{Internet Research}, 32\penalty0 (7):\penalty0 1--31, May 2022.
\newblock \doi{10.1108/intr-08-2021-0600}.
\newblock URL \url{https://doi.org/10.1108/intr-08-2021-0600}.

\bibitem[Lakkaraju \& Bastani(2020)Lakkaraju and
  Bastani]{10.1145/3375627.3375833}
Lakkaraju, H. and Bastani, O.
\newblock "how do i fool you?": Manipulating user trust via misleading black
  box explanations.
\newblock In \emph{Proceedings of the AAAI/ACM Conference on AI, Ethics, and
  Society}, AIES '20, pp.\  79–85, New York, NY, USA, 2020. Association for
  Computing Machinery.
\newblock ISBN 9781450371100.
\newblock \doi{10.1145/3375627.3375833}.
\newblock URL \url{https://doi.org/10.1145/3375627.3375833}.

\bibitem[Li et~al.(2018)Li, Liu, Chen, and Rudin]{Lia}
Li, O., Liu, H., Chen, C., and Rudin, C.
\newblock Deep learning for case-based reasoning through prototypes: A neural
  network that explains its predictions.
\newblock In \emph{AAAI}, 2018.

\bibitem[Liao \& Varshney(2021)Liao and
  Varshney]{DBLP:journals/corr/abs-2110-10790}
Liao, Q.~V. and Varshney, K.~R.
\newblock Human-centered explainable {AI} {(XAI):} from algorithms to user
  experiences.
\newblock \emph{CoRR}, abs/2110.10790, 2021.
\newblock URL \url{https://arxiv.org/abs/2110.10790}.

\bibitem[Lim et~al.(2019)Lim, Yang, Abdul, and Wang]{Lim2019}
Lim, B.~Y., Yang, Q., Abdul, A., and Wang, D.
\newblock {Why these Explanations? Selecting Intelligibility Types for
  Explanation Goals}.
\newblock pp.\ ~7, 2019.
\newblock \doi{10.1145/1234567890}.
\newblock URL \url{https://doi.org/10.1145/1234567890}.

\bibitem[Lundberg \& Lee(2017)Lundberg and Lee]{NIPS2017_8a20a862}
Lundberg, S.~M. and Lee, S.-I.
\newblock A unified approach to interpreting model predictions.
\newblock In Guyon, I., Luxburg, U.~V., Bengio, S., Wallach, H., Fergus, R.,
  Vishwanathan, S., and Garnett, R. (eds.), \emph{Advances in Neural
  Information Processing Systems}, volume~30. Curran Associates, Inc., 2017.
\newblock URL
  \url{https://proceedings.neurips.cc/paper/2017/file/8a20a8621978632d76c43dfd28b67767-Paper.pdf}.

\bibitem[Lundberg et~al.(2017)Lundberg, Allen, and Lee]{Lundberg}
Lundberg, S.~M., Allen, P.~G., and Lee, S.-I.
\newblock {A Unified Approach to Interpreting Model Predictions}.
\newblock In \emph{Advances in Neural Information Processing Systems 30}, pp.\
  4765--4774, 2017.
\newblock URL \url{https://github.com/slundberg/shap}.

\bibitem[Mahendran \& Vedaldi(2014)Mahendran and Vedaldi]{Mahendran2014a}
Mahendran, A. and Vedaldi, A.
\newblock {Understanding Deep Image Representations by Inverting Them}.
\newblock nov 2014.
\newblock URL \url{http://arxiv.org/abs/1412.0035}.

\bibitem[Mehrabi et~al.(2019)Mehrabi, Morstatter, Saxena, Lerman, and
  Galstyan]{Mehrabi2019}
Mehrabi, N., Morstatter, F., Saxena, N., Lerman, K., and Galstyan, A.
\newblock {A Survey on Bias and Fairness in Machine Learning}.
\newblock 2019.
\newblock URL \url{http://arxiv.org/abs/1908.09635}.

\bibitem[Miller(2019)]{Miller2017}
Miller, T.
\newblock Explanation in artificial intelligence: Insights from the social
  sciences.
\newblock \emph{Artificial Intelligence}, 267:\penalty0 1--38, 2019.
\newblock ISSN 0004-3702.
\newblock \doi{https://doi.org/10.1016/j.artint.2018.07.007}.
\newblock URL
  \url{https://www.sciencedirect.com/science/article/pii/S0004370218305988}.

\bibitem[Miller et~al.(2017)Miller, Hower, and Sonenberg]{Miller}
Miller, T., Hower, P., and Sonenberg, L.
\newblock {Explainable AI: beware of inmates running the asylum}.
\newblock In \emph{IJCAI 2017 workshop on explainable artificial intelligence
  (XAI)}, number October, pp.\  363, 2017.
\newblock \doi{10.1016/j.foodchem.2017.11.091}.
\newblock URL \url{http://home.earthlink.net/}.

\bibitem[Ming et~al.(2019)Ming, Qu, and Bertini]{Ming2019}
Ming, Y., Qu, H., and Bertini, E.
\newblock {RuleMatrix: Visualizing and Understanding Classifiers with Rules}.
\newblock \emph{IEEE Transactions on Visualization and Computer Graphics},
  25\penalty0 (1):\penalty0 342--352, jan 2019.
\newblock ISSN 1077-2626.
\newblock \doi{10.1109/TVCG.2018.2864812}.
\newblock URL \url{https://ieeexplore.ieee.org/document/8440085/}.

\bibitem[Mohseni et~al.(2021)Mohseni, Zarei, and Ragan]{10.1145/3387166}
Mohseni, S., Zarei, N., and Ragan, E.~D.
\newblock A multidisciplinary survey and framework for design and evaluation of
  explainable ai systems.
\newblock \emph{ACM Trans. Interact. Intell. Syst.}, 11\penalty0 (3–4), sep
  2021.
\newblock ISSN 2160-6455.
\newblock \doi{10.1145/3387166}.
\newblock URL \url{https://doi.org/10.1145/3387166}.

\bibitem[Narayanan et~al.(2018)Narayanan, Chen, He, Kim, Gershman, and
  Doshi-Velez]{Narayanan2018}
Narayanan, M., Chen, E., He, J., Kim, B., Gershman, S., and Doshi-Velez, F.
\newblock {How do Humans Understand Explanations from Machine Learning Systems?
  An Evaluation of the Human-Interpretability of Explanation}.
\newblock feb 2018.
\newblock URL \url{http://arxiv.org/abs/1802.00682}.

\bibitem[Nguyen et~al.(2021)Nguyen, Kim, and Nguyen]{NEURIPS2021_de043a5e}
Nguyen, G., Kim, D., and Nguyen, A.
\newblock The effectiveness of feature attribution methods and its correlation
  with automatic evaluation scores.
\newblock In Ranzato, M., Beygelzimer, A., Dauphin, Y., Liang, P., and Vaughan,
  J.~W. (eds.), \emph{Advances in Neural Information Processing Systems},
  volume~34, pp.\  26422--26436. Curran Associates, Inc., 2021.
\newblock URL
  \url{https://proceedings.neurips.cc/paper/2021/file/de043a5e421240eb846da8effe472ff1-Paper.pdf}.

\bibitem[Nunes \& Jannach(2017)Nunes and Jannach]{Nunes2017}
Nunes, I. and Jannach, D.
\newblock {A systematic review and taxonomy of explanations in decision support
  and recommender systems}.
\newblock \emph{User Modeling and User-Adapted Interaction}, 27\penalty0
  (3-5):\penalty0 393--444, 2017.
\newblock ISSN 15731391.
\newblock \doi{10.1007/s11257-017-9195-0}.

\bibitem[Olah et~al.(2017)Olah, Mordvintsev, and Schubert]{Olah2017}
Olah, C., Mordvintsev, A., and Schubert, L.
\newblock Feature visualization.
\newblock \emph{Distill}, 2\penalty0 (11), November 2017.
\newblock \doi{10.23915/distill.00007}.
\newblock URL \url{https://doi.org/10.23915/distill.00007}.

\bibitem[Ribeiro et~al.(2016)Ribeiro, Singh, and Guestrin]{Ribeiro2016a}
Ribeiro, M.~T., Singh, S., and Guestrin, C.
\newblock "why should i trust you?": Explaining the predictions of any
  classifier.
\newblock In \emph{Proceedings of the 22nd ACM SIGKDD International Conference
  on Knowledge Discovery and Data Mining}, KDD '16, pp.\  1135–1144, New
  York, NY, USA, 2016. Association for Computing Machinery.
\newblock ISBN 9781450342322.
\newblock \doi{10.1145/2939672.2939778}.
\newblock URL \url{https://doi.org/10.1145/2939672.2939778}.

\bibitem[Ribeiro et~al.(2018)Ribeiro, Singh, and Guestrin]{Ribeiro}
Ribeiro, M.~T., Singh, S., and Guestrin, C.
\newblock Anchors: High-precision model-agnostic explanations.
\newblock In \emph{AAAI Conference on Artificial Intelligence (AAAI)}, 2018.

\bibitem[Rosson \& Carroll(2002)Rosson and Carroll]{10.5555/772072.772137}
Rosson, M.~B. and Carroll, J.~M.
\newblock \emph{Scenario-Based Design}, pp.\  1032–1050.
\newblock L. Erlbaum Associates Inc., USA, 2002.
\newblock ISBN 0805838384.

\bibitem[Schneider \& Handali(2019)Schneider and
  Handali]{Schneider2019PersonalizedEI}
Schneider, J. and Handali, J.~P.
\newblock Personalized explanation in machine learning.
\newblock \emph{ArXiv}, abs/1901.00770, 2019.

\bibitem[Shang et~al.(2022)Shang, Feng, and Shah]{10.1145/3531146.3533189}
Shang, R., Feng, K. J.~K., and Shah, C.
\newblock Why am i not seeing it? understanding users’ needs for
  counterfactual explanations in everyday recommendations.
\newblock In \emph{2022 ACM Conference on Fairness, Accountability, and
  Transparency}, FAccT '22, pp.\  1330–1340, New York, NY, USA, 2022.
  Association for Computing Machinery.
\newblock \doi{10.1145/3531146.3533189}.
\newblock URL \url{https://doi.org/10.1145/3531146.3533189}.

\bibitem[Shapley(1951)]{RM-670-PR}
Shapley, L.~S.
\newblock \emph{Notes on the n-Person Game -- II: The Value of an n-Person
  Game}.
\newblock RAND Corporation, Santa Monica, CA, 1951.

\bibitem[Shen \& Huang(2020)Shen and Huang]{shen2020useful}
Shen, H. and Huang, T.-H.
\newblock How useful are the machine-generated interpretations to general
  users? a human evaluation on guessing the incorrectly predicted labels.
\newblock In \emph{Proceedings of the AAAI Conference on Human Computation and
  Crowdsourcing}, volume~8, pp.\  168--172, 2020.

\bibitem[Simonyan et~al.(2013)Simonyan, Vedaldi, and Zisserman]{Simonyan}
Simonyan, K., Vedaldi, A., and Zisserman, A.
\newblock {Deep Inside Convolutional Networks: Visualising Image Classification
  Models and Saliency Maps}.
\newblock Technical report, 2013.
\newblock URL \url{https://arxiv.org/abs/1312.6034v2}.

\bibitem[Siu et~al.(2021)Siu, Pena, Chen, Zhou, Lopez, Palko, Chang, and
  Allen]{siu2021evaluation}
Siu, H.~C., Pena, J.~D., Chen, E., Zhou, Y., Lopez, V., Palko, K., Chang,
  K.~C., and Allen, R.~E.
\newblock Evaluation of human-{AI} teams for learned and rule-based agents in
  hanabi.
\newblock In Beygelzimer, A., Dauphin, Y., Liang, P., and Vaughan, J.~W.
  (eds.), \emph{Advances in Neural Information Processing Systems}, 2021.
\newblock URL \url{https://openreview.net/forum?id=bB-l0cnS3E}.

\bibitem[Suresh et~al.(2021)Suresh, Gomez, Nam, and
  Satyanarayan]{10.1145/3411764.3445088}
Suresh, H., Gomez, S.~R., Nam, K.~K., and Satyanarayan, A.
\newblock Beyond expertise and roles: A framework to characterize the
  stakeholders of interpretable machine learning and their needs.
\newblock In \emph{Proceedings of the 2021 CHI Conference on Human Factors in
  Computing Systems}, CHI '21, New York, NY, USA, 2021. Association for
  Computing Machinery.
\newblock ISBN 9781450380966.
\newblock \doi{10.1145/3411764.3445088}.
\newblock URL \url{https://doi.org/10.1145/3411764.3445088}.

\bibitem[Taesiri et~al.(2022)Taesiri, Nguyen, and Nguyen]{taesiri2022visual}
Taesiri, M.~R., Nguyen, G., and Nguyen, A.
\newblock Visual correspondence-based explanations improve {AI} robustness and
  human-{AI} team accuracy.
\newblock In Oh, A.~H., Agarwal, A., Belgrave, D., and Cho, K. (eds.),
  \emph{Advances in Neural Information Processing Systems}, 2022.
\newblock URL \url{https://openreview.net/forum?id=UavQ9HYye6n}.

\bibitem[Tan et~al.(2018)Tan, Caruana, Hooker, Koch, and Gordo]{Tan}
Tan, S., Caruana, R., Hooker, G., Koch, P., and Gordo, A.
\newblock {Learning Global Additive Explanations for Neural Nets Using Model
  Distillation}.
\newblock Technical report, 2018.
\newblock URL \url{https://youtu.be/ErQYwNqzEdc.}

\bibitem[Turner et~al.(2020)Turner, Kaushik, Huang, and Varanasi]{Turner}
Turner, A., Kaushik, M., Huang, M.-T., and Varanasi, S.
\newblock {Calibrating Trust in AI-Assisted Decision Making}.
\newblock 2020.

\bibitem[Ustun \& Rudin(2016)Ustun and Rudin]{DBLP:journals/ml/UstunR16}
Ustun, B. and Rudin, C.
\newblock Supersparse linear integer models for optimized medical scoring
  systems.
\newblock \emph{Mach. Learn.}, 102\penalty0 (3):\penalty0 349--391, 2016.
\newblock URL \url{https://doi.org/10.1007/s10994-015-5528-6}.

\bibitem[Vaughan \& Wallach(2021)Vaughan and
  Wallach]{10.7551/mitpress/12186.003.0014}
Vaughan, J.~W. and Wallach, H.
\newblock {A Human-Centered Agenda for Intelligible Machine Learning}.
\newblock In \emph{{Machines We Trust: Perspectives on Dependable AI}}. The MIT
  Press, 08 2021.
\newblock ISBN 9780262366212.
\newblock \doi{10.7551/mitpress/12186.003.0014}.
\newblock URL \url{https://doi.org/10.7551/mitpress/12186.003.0014}.

\bibitem[Wolf(2019)]{Wolf2019}
Wolf, C.~T.
\newblock {Explainability scenarios: Towards scenario-based XAI design}.
\newblock \emph{International Conference on Intelligent User Interfaces,
  Proceedings IUI}, Part F1476:\penalty0 252--257, 2019.
\newblock \doi{10.1145/3301275.3302317}.

\bibitem[Yang et~al.(2020)Yang, Huang, Scholtz, and
  Arendt]{10.1145/3377325.3377480}
Yang, F., Huang, Z., Scholtz, J., and Arendt, D.~L.
\newblock How do visual explanations foster end users' appropriate trust in
  machine learning?
\newblock In \emph{Proceedings of the 25th International Conference on
  Intelligent User Interfaces}, IUI '20, pp.\  189–201, New York, NY, USA,
  2020. Association for Computing Machinery.
\newblock ISBN 9781450371186.
\newblock \doi{10.1145/3377325.3377480}.
\newblock URL \url{https://doi.org/10.1145/3377325.3377480}.

\bibitem[Yang et~al.(2017)Yang, Rudin, and Seltzer]{Yang2017}
Yang, H., Rudin, C., and Seltzer, M.
\newblock Scalable bayesian rule lists.
\newblock In \emph{Proceedings of the 34th International Conference on Machine
  Learning - Volume 70}, ICML'17, pp.\  3921–3930. JMLR.org, 2017.

\bibitem[Yin et~al.(2019)Yin, Wortman~Vaughan, and Wallach]{Yin2019}
Yin, M., Wortman~Vaughan, J., and Wallach, H.
\newblock Understanding the effect of accuracy on trust in machine learning
  models.
\newblock In \emph{Proceedings of the 2019 CHI Conference on Human Factors in
  Computing Systems}, CHI '19, pp.\  1–12, New York, NY, USA, 2019.
  Association for Computing Machinery.
\newblock ISBN 9781450359702.
\newblock \doi{10.1145/3290605.3300509}.
\newblock URL \url{https://doi.org/10.1145/3290605.3300509}.

\bibitem[Zhang et~al.(2018)Zhang, Yang, Ma, and Wu]{Zhang2018d}
Zhang, Q., Yang, Y., Ma, H., and Wu, Y.~N.
\newblock {Interpreting CNNs via Decision Trees}.
\newblock jan 2018.
\newblock URL \url{http://arxiv.org/abs/1802.00121}.

\bibitem[Zhang et~al.(2020{\natexlab{a}})Zhang, Liao, and Bellamy]{Zhang2020a}
Zhang, Y., Liao, Q.~V., and Bellamy, R. K.~E.
\newblock {Effect of Confidence and Explanation on Accuracy and Trust
  Calibration in AI-Assisted Decision Making}.
\newblock \emph{FAT* 2020 - Proceedings of the 2020 Conference on Fairness,
  Accountability, and Transparency}, pp.\  295--305, jan 2020{\natexlab{a}}.
\newblock \doi{10.1145/3351095.3372852}.
\newblock URL \url{http://arxiv.org/abs/2001.02114
  http://dx.doi.org/10.1145/3351095.3372852}.

\bibitem[Zhang et~al.(2020{\natexlab{b}})Zhang, {Vera Liao}, and
  Bellamy]{Zhang2020}
Zhang, Y., {Vera Liao}, Q., and Bellamy, R.~K.
\newblock {Efect of confidence and explanation on accuracy and trust
  calibration in AI-assisted decision making}.
\newblock \emph{FAT* 2020 - Proceedings of the 2020 Conference on Fairness,
  Accountability, and Transparency}, pp.\  295--305, 2020{\natexlab{b}}.
\newblock \doi{10.1145/3351095.3372852}.

\bibitem[Zhou et~al.(2015)Zhou, Khosla, Lapedriza, Oliva, and Torralba]{Zhoub}
Zhou, B., Khosla, A., Lapedriza, A., Oliva, A., and Torralba, A.
\newblock {Learning Deep Features for Discriminative Localization}.
\newblock Technical report, 2015.
\newblock URL \url{http://cnnlocalization.csail.mit.edu}.

\end{thebibliography}

\end{document}